\documentclass[10pt,conference]{IEEEtran}

\IEEEoverridecommandlockouts
\usepackage{adjustbox}
\usepackage{longtable}
\usepackage{cite}
\usepackage{underscore}
\usepackage{amsmath,amssymb,amsfonts}
\usepackage{algorithmic}
\usepackage{graphicx}
\usepackage{textcomp}
\usepackage{xcolor}
\usepackage[euler]{textgreek}
\usepackage{color}
\usepackage{url}
\usepackage{multirow}
\usepackage{tabularx}
\usepackage{subfig}
\usepackage{enumitem}
\usepackage{caption}
\usepackage{pdfcomment}

\usepackage{afterpage}
\newcommand\headerpage{
    \thispagestyle{empty}
    \renewcommand\ttdefault{cmvtt}

    \begin{figure*}
        \centering
        {\huge \texttt{NEUKONFIG}: Reducing Edge Service Downtime When Repartitioning DNNs\\} \vspace{0.5cm}
        
        {\large Ayesha Abdul Majeed, Peter Kilpatrick, Ivor Spence, and Blesson Varghese\\
        School of Electronics, Electrical Engineering and Computer Science\\Queen's University Belfast, UK\\
        E-mail: \{aabdulmajeed01, p.kilpatrick, i.spence, b.varghese\}@qub.ac.uk} \vspace{12cm}
        
        {\large Accepted to the 9th IEEE International Conference on Cloud Engineering, 2021 \\
        Document version: Pre-print} \vspace{2cm}
        
        {\large ©2021 IEEE. Personal use of this material is permitted. Permission from IEEE must be obtained for all other uses, in any current or future media, including reprinting/republishing this material for advertising or promotional purposes, creating new collective works, for resale or redistribution to servers or lists, or reuse of any copyrighted component of this work in other works.}

    \end{figure*}

    \addtocounter{page}{-1}
    \newpage
    }

\begin{document}

\renewcommand\ttdefault{cmvtt}

\headerpage

\title{\texttt{NEUKONFIG}: Reducing Edge Service Downtime When Repartitioning DNNs}


\author{\IEEEauthorblockN{Ayesha Abdul Majeed, Peter Kilpatrick, Ivor Spence, and Blesson Varghese}
\IEEEauthorblockA{\textit{School of Electronics, Electrical Engineering and Computer Science, Queen's University Belfast, UK}\\
E-mail: \{aabdulmajeed01, p.kilpatrick, i.spence, b.varghese\}@qub.ac.uk}
}

\maketitle
\thispagestyle{plain}
\pagestyle{plain}

\begin{abstract}

Deep Neural Networks (DNNs) may be partitioned across the edge and the cloud to improve the performance efficiency of inference. DNN partitions are determined based on operational conditions such as network speed. When operational conditions change DNNs will need to be repartitioned to maintain the overall performance. However, repartitioning using existing approaches, such as Pause and Resume, will incur a service downtime on the edge. This paper presents the \texttt{NEUKONFIG} framework that identifies the service downtime incurred when repartitioning DNNs and proposes approaches for reducing edge service downtime. The proposed approaches are based on `\textit{Dynamic Switching}' in which, when the network speed changes and given an existing edge-cloud pipeline, a new edge-cloud pipeline is initialised with new DNN partitions. Incoming inference requests are switched to the new pipeline for processing data. Two dynamic switching scenarios are considered: when a second edge-cloud pipeline is always running and when a second pipeline is only initialised when the network speed changes. Experimental studies are carried out on a lab-based testbed to demonstrate that Dynamic Switching reduces the downtime by at least an order of magnitude when compared to a baseline using Pause and Resume that has a downtime of 6 seconds. A trade-off in the edge service downtime and memory required is noted. The Dynamic Switching approach that requires the same amount of memory as the baseline reduces the edge service downtime to 0.6 seconds and to less than 1 millisecond in the best case when twice the amount of memory as the baseline is available.

\end{abstract}


\section{Introduction}
\label{sec:introduction}
\begin{figure}[t]
    \centering
    \includegraphics[width=0.49\textwidth]{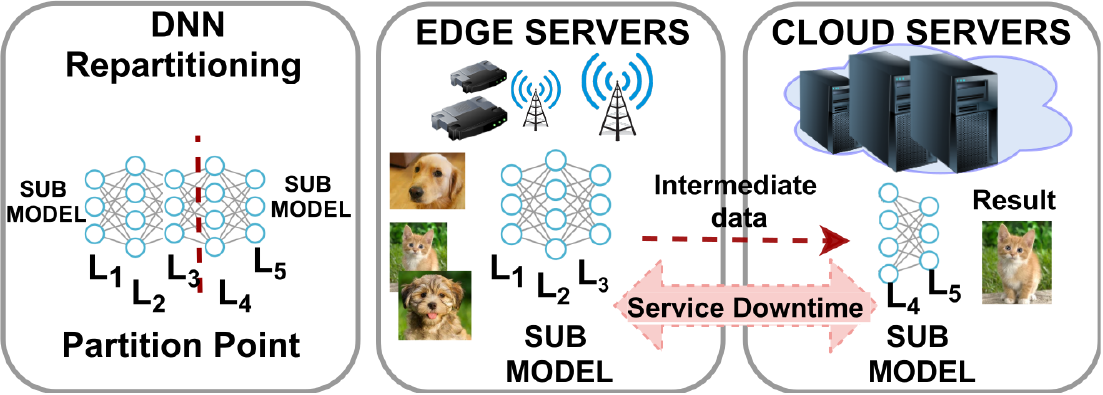}
    \caption{A repartitioning of DNN model is required due to system load or unstable network once a model has been deployed on Edge server.}
    \label{fig:dnn}
\end{figure}

Deep Neural Networks (DNNs) are widely adopted by Internet applications to process input data, such as speech, image and video, for label classification, pattern recognition or object detection~\cite{dnn-01, dnn-02}. 
The need of applications to minimise upstream network traffic, reduce end-to-end latency, and enhance data privacy has resulted in DNN inference being performed outside cloud data centres where they are typically performed; this paradigm is referred to as edge computing~\cite{intro-01, intro-02, intro-03}.  
An ideal location for DNN inference is therefore on resources located at the edge of the network, such as home routers, gateways or micro data centres~\cite{lockhart2020scission, zhou2019edge, wang2020convergence}. 
The key challenge is that edge resources for running large DNNs that require significant processing and memory are computationally constrained when compared to the cloud. 
More recently, this challenge has been mitigated by partitioning DNNs across the edge and the cloud for accelerating inference\cite{colloborative2019, dynamicDNN,wang2019adda, neurosurgeon}. 
This is achieved by dividing a single DNN model into multiple partitions by exploiting the essential characteristic of a DNN, which is that it is composed of a sequence of layers.
Each partition contains a sequence of layers. 
The first partition is processed on the edge and the intermediate result is transferred via the network and provided as input to the second partition on the cloud (shown in Figure~\ref{fig:dnn}).
At which layer a DNN should be divided is based on the operational conditions of the edge-cloud environment (state of the network, CPU on the edge etc)~\cite{mcnamee2020case, lockhart2020scission}. 

Current research that develops DNN partitioning approaches only accounts for the initial mapping of DNN partitions to the edge-cloud environment~\cite{dynamicDNN, wang2019adda}, and does not consider repartitioning. 
In this paper, it is hypothesised that when operational conditions change the DNN will need to be repartitioned.
Thus, a different sequence of layers will need to be assigned to the edge and cloud. 

It is also hypothesised that repartitioning will incur a service downtime on the edge. This is because existing approaches pause the running application and then resume with the new partitions~\cite{da2019multi, ma2018efficient}.

Pausing and resuming is not a feasible approach for latency-critical applications that have real-time constraints. 
Consider for example, if DNNs had to be repartitioned in the face of variable network conditions for supporting the navigation of a robot on a factory floor occupied by humans. 
A service downtime of even a few seconds on the edge while partitioning DNNs could make the robot less safe for humans.

Therefore, there is a need for \textit{reducing edge service downtime when repartitioning DNNs}, which is the focus of this paper. The above hypotheses will be verified and an approach for reducing edge service downtime is developed. 
More specifically, the following three questions are considered: 

\textbf{(Q1)} \textit{When is DNN repartitioning required?}
Based on empirical results obtained by examining well-known production DNNs, such as VGG-19~\cite{vgg} and MobileNetV2~\cite{mobilenetv2}, scenarios in which a DNN is to be partitioned are observed. It is noted that variation in network speed is a key contributing factor to when DNN repartitioning is required.  

\textbf{(Q2)} \textit{What is the edge service downtime associated with repartitioning a DNN?}
This question is addressed by developing a baseline approach (motivated by a Pause and Resume approach presented in the literature for streaming applications~\cite{da2019multi}) to facilitate DNN repartitioning. The baseline `pauses' the running DNN, repartitions the DNN and `resumes' execution. The Pause and Resume approach identifies the downtime on the edge server when repartitioning DNNs. 

\textbf{(Q3)} \textit{How can edge service downtime be reduced while repartitioning a DNN?}
The downtime when repartitioning a DNN is reduced by developing a `Dynamic Switching' approach that aims to enhance the resilience of an edge server for DNN-based applications. For this, given an edge-cloud pipeline to execute DNN partitions, when repartitioning needs to occur, a new edge-cloud pipeline is instantiated while retaining the initial pipeline until it is no longer required. 


The above questions are addressed in the context of an experimental framework named as \texttt{NEUKONFIG}. This framework: (i) identifies the scenarios in which a DNN needs to be repartitioned, (ii) incorporates the Pause and Resume approach for repartitioning DNNs that is used as a baseline, (iii) develops the Dynamic Switching approach for reducing edge service downtime, and (iv) gathers empirical data from an experimental edge-cloud test-bed to guide the observations of this paper.

Two scenarios with different resource requirements are considered to optimise the Dynamic Switching approach: (i) a second edge-cloud pipeline is always running, and (ii) a second pipeline is initialised only when the network speed changes.
For example, the Pause and Resume approach requires over 6 seconds for repartitioning a DNN when the network speed drops from 20Mbps to 5Mbps (and vice-versa). However, this time can be brought down to under 1 millisecond when a second edge-cloud pipeline is always running and to 0.6 seconds when the second edge-cloud pipeline is initialised only when the network speed changes.
A trade-off with edge service downtime is observed in that additional memory resources are required to initialise or maintain a second edge-cloud pipeline. 
For memory resources similar to that required by the Pause and Resume approach, the Dynamic Switching approach significantly reduces the edge service downtime by more than an order of magnitude. 


This paper makes the following contributions: (i) Quantifies the edge service downtime for repartitioning DNNs and thus identifies the limitation of using approaches reported in the literature. (ii) Develops `Dynamic Switching', a novel approach to reducing edge service downtime and quantifies its impact for seamlessly repartitioning DNNs. (iii) Demonstrates that Dynamic Switching can reduce the edge service downtime during DNN repartitioning. 

The remainder of this paper is organized as follows: Section~\ref{sec:background} identifies the scenarios when a DNN should be repartitioned by examining production DNNs to address Q1 above. Section~\ref{sec:technique} initially presents a naive approach for partitioning DNNs using a Pause and Resume approach that enables us to identify the edge service downtime when repartitioning (addresses Q2). Subsequently, the Dynamic Switching approach is presented that addresses Q3. Section~\ref{sec:experiments} presents the implementation and experimental results obtained. Section~\ref{sec:relatedwork} presents related work. Section~\ref{sec:conclusions} concludes the paper by considering the limitations of the current work and the avenues for future research.

\section{Identifying Scenarios for Repartitioning DNNs}
\label{sec:background}
This section presents the scenarios that are identified by \texttt{NEUKONFIG} for DNN repartitioning (addresses Q1 raised in Section~\ref{sec:introduction}). Two popular pre-trained production DNNs are considered, VGG-19~\cite{vgg} and MobileNetV2~\cite{mobilenetv2}. 

\subsection{Background}

In this paper, partitioning a DNN is a reference to splitting a DNN model at a suitable layer into partitions and distributing them over the edge and the cloud. The first partition is processed at the edge and the intermediate result is transferred via network and passed as input to the second partition that executes on the cloud (shown in Figure~\ref{fig:dnn}). The total inference time (end-to-end latency) for
such a partitioned DNN is defined in Equation~\ref{eqn:end-to-end}:
\begin{equation}
T_{inf}=T_{e}+T_{t}+T_{c} 
\label{eqn:end-to-end}
\end{equation}
where $T_{e}$ is the execution time on the edge of the first partition, $T_{t}$ is the time to transfer the intermediate data from the first partition to the cloud server, and $T_{c}$ is the execution time of the second partition on the cloud. 

The ideal partitioning point of a DNN is identified by accounting for the trade-off between the time taken for computation and the time taken to transfer the intermediate data. Partitioning at different points of a DNN will result in different times for computation on the cloud and the edge and a different time to transfer data between partitions (volume of data transferred between layers is different), thereby affecting the overall end-to-end latency.

The performance of an edge resource varies over time due to the variation in resource availability and network between the edge and cloud. This is expected to affect the performance of DNN applications that are deployed on these resources. Therefore, the DNN must adapt to changes in the operational environment. A new partitioning point of the DNN must be identified to this end (referred to as DNN reparitioning), so that the end-to-end latency can be minimised when compared to that of the initial partitions. 

Consider for example if a DNN with five layers was initially partitioned at $L_3$, such that $L_1$ - $L_3$ execute on the edge and $L_4$ - $L_5$ execute on the cloud. The partitioning point may have to change from $L_3$ to $L_2$ for maintaining the required level of performance for the DNN (as shown in Figure~\ref{fig:dnn}). The aim of this section is to identify the scenarios in which DNN repartitioning would be required. 

An experimental environment consisting of a cloud virtual machine (VM) running Ubuntu 18.10 with 8 virtual CPUs and 8GB RAM and an edge VM running Ubuntu 18.10 with 2 virtual CPUs and 4GB RAM was used.
Docker 18.09-ce is installed on both VMs.
The outbound network traffic from the edge to the cloud is emulated using the Linux Traffic Control (\texttt{tc})\footnote{https://linux.die.net/man/8/tc} tool.
Two types of pre-trained DNNs, namely sequential (VGG-19) and non-sequential (MobileNetV2) are considered on the Keras\footnote{https://keras.io/api/applications/} platform. 

The layers of the DNN are profiled to gather empirically the computation time of each layer on the edge and cloud, the size of data transferred between layers at the split point, and the time taken to transfer this data from the edge to the cloud. This is straightforward for sequential DNNs. However, for non-sequential DNNs the parallel paths need to be accounted for. In this paper, layers in the parallel path of a non-sequential DNN are not partitioned. Rather each parallel region is treated as a block (similar approaches are used in the literature~\cite{lockhart2020scission}). 

\begin{figure}[t]
\begin{center}
	\subfloat[Latency when the network speed is 20Mbps]
	{\label{fig:vgg-20}
	\includegraphics[width=0.48\textwidth]
	{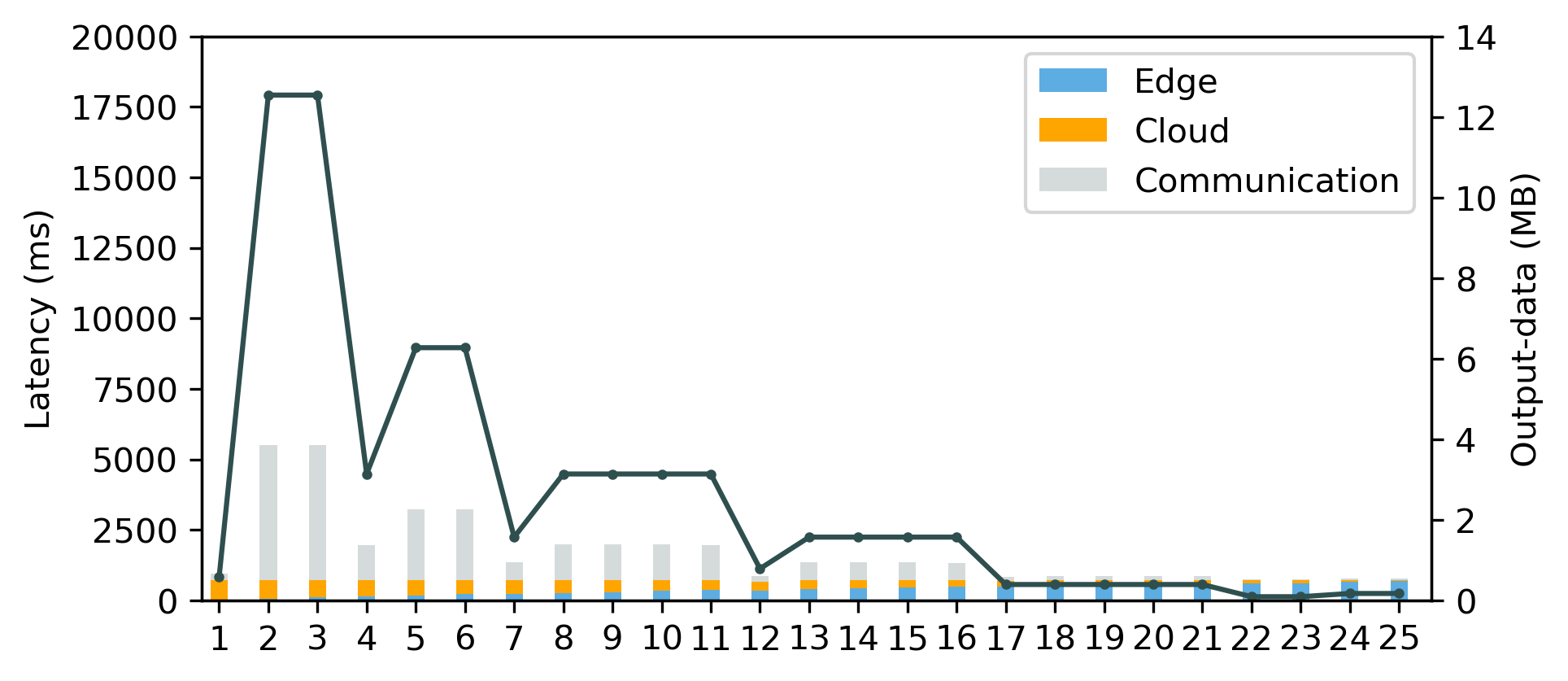}} 
	
	\subfloat[Latency when the network speed drops to 5Mbps]
	{\label{fig:vgg-5}
	\includegraphics[width=0.48\textwidth]
	{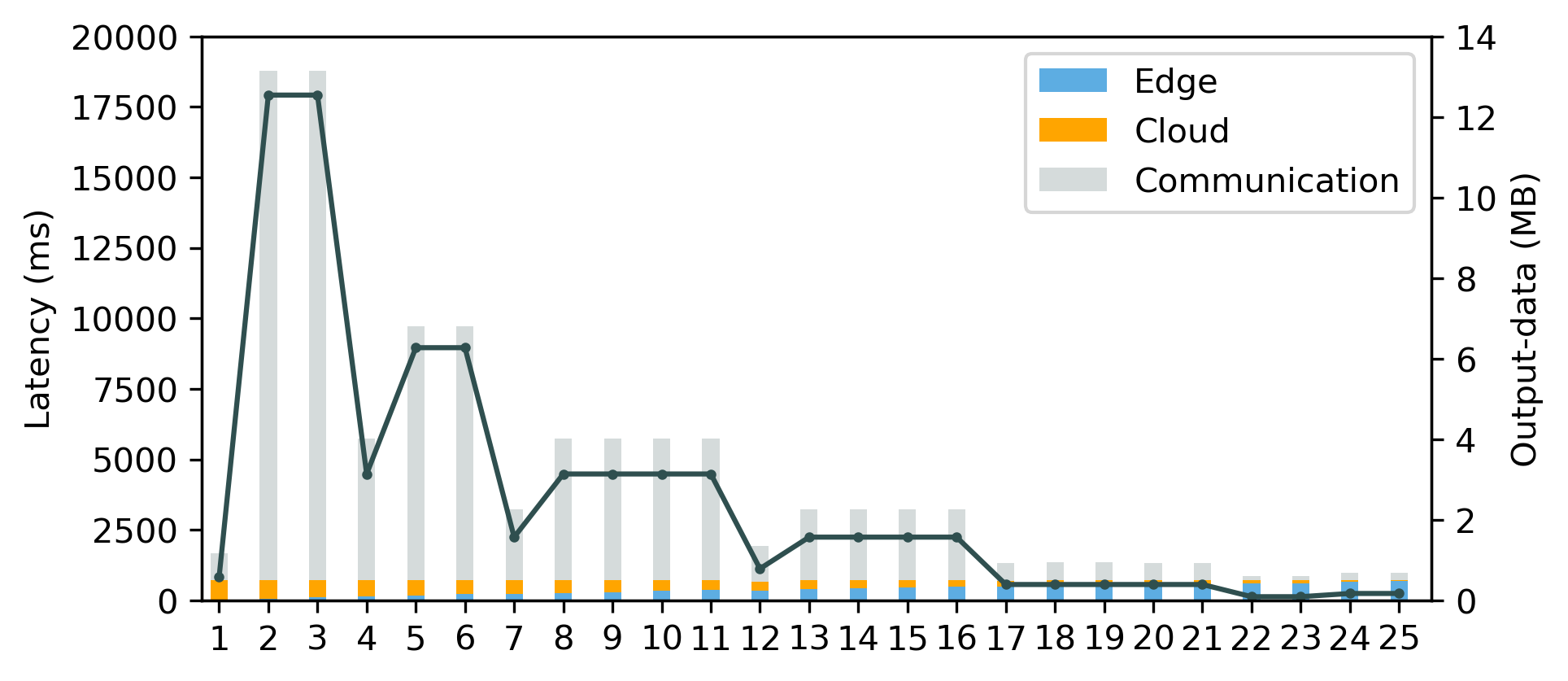}}
\end{center}
\caption{End-to-end latency and the size of output data transferred between the edge and the cloud for VGG-19 for different partition points.}
\label{fig:vgg-comm}
\end{figure}

\begin{figure}[t]
\begin{center}
	\subfloat[Latency when the network speed is 20Mbps]
	{\label{fig:mobilenet-20}
	\includegraphics[width=0.48\textwidth]
	{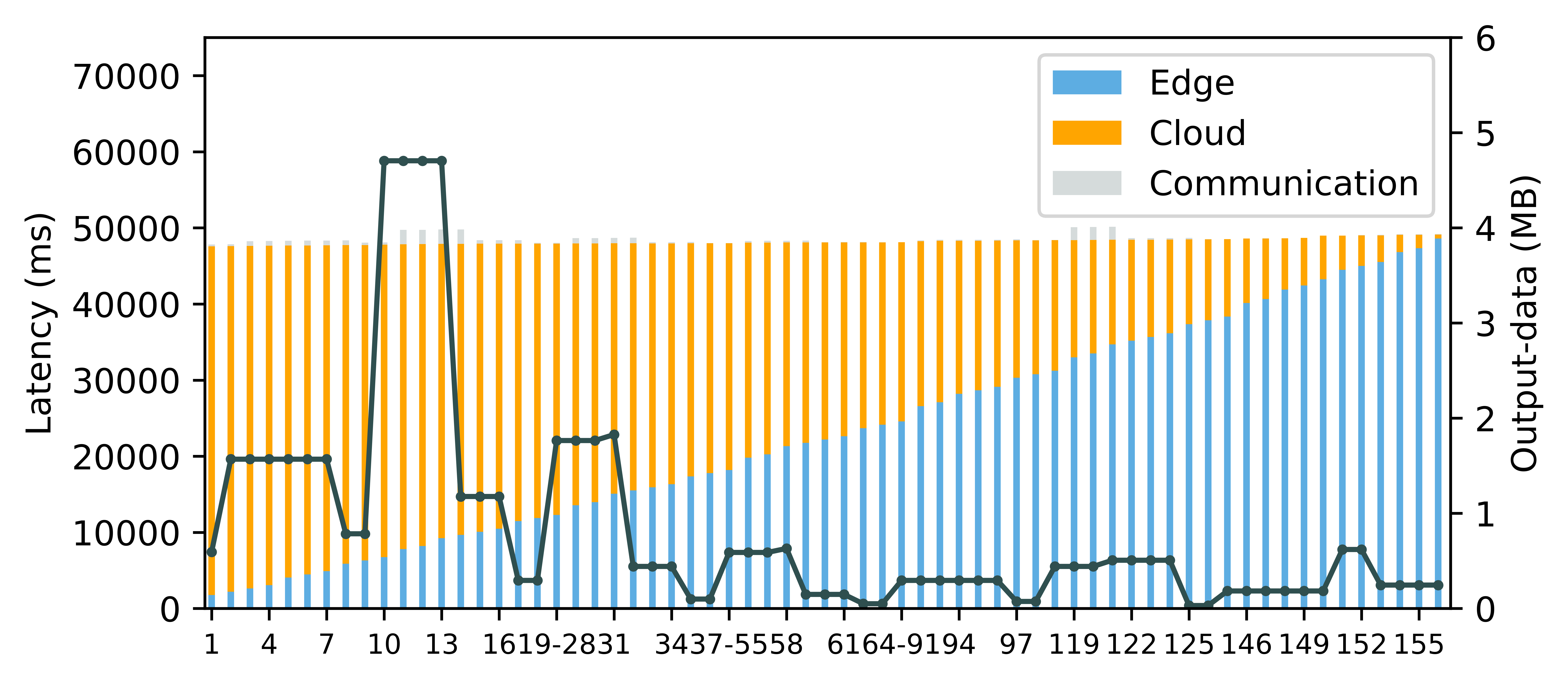}}
	
	\subfloat[Latency when the network speed drops to 5Mbps]
	{\label{fig:mobilenet-5}
	\includegraphics[width=0.48\textwidth]
	{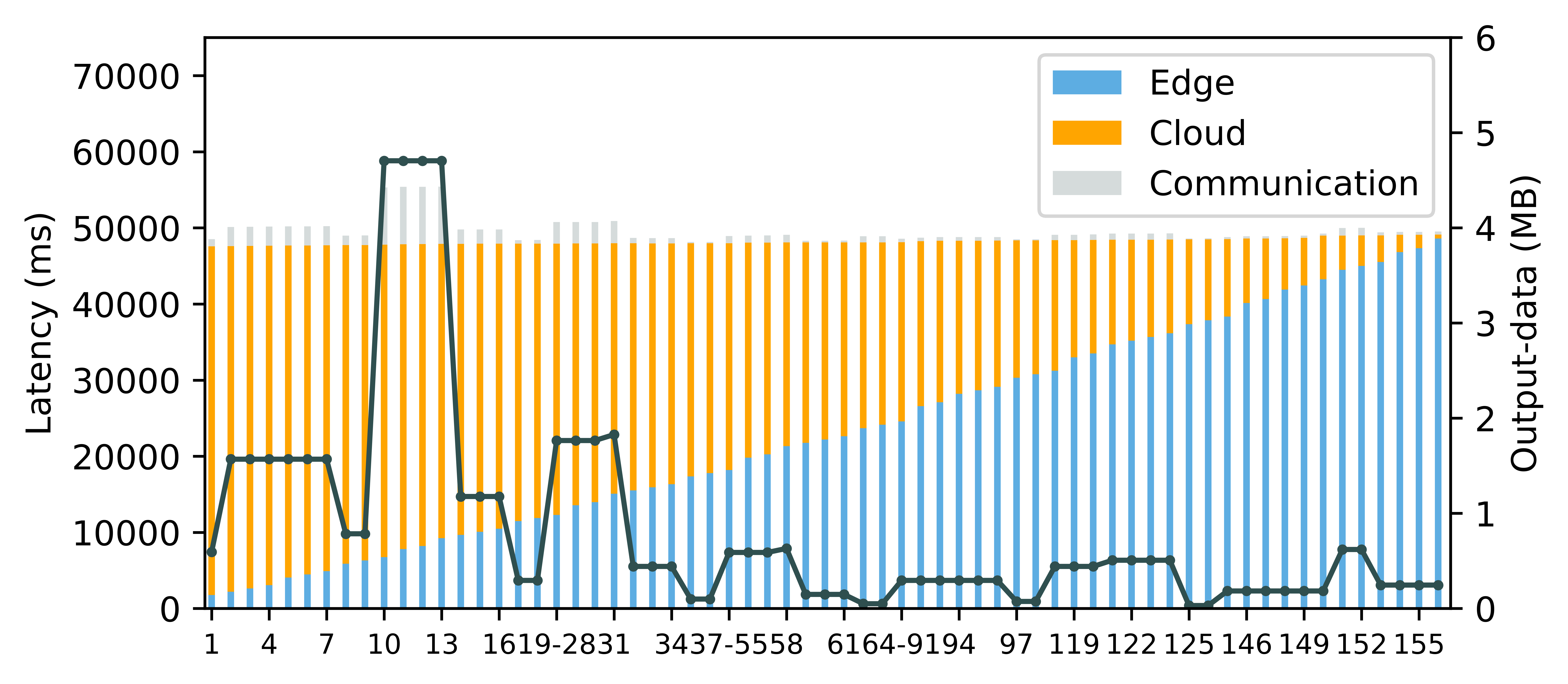}}
\end{center}
\caption{End-to-end latency and the size of output data transferred between the edge and the cloud for MobileNetV2 for different partition points.}
\label{fig:mobilenetv2-comm}
\end{figure}

\subsection{Scenarios for Repartitioning}

The above experimental set up was used to identify scenarios in which DNN repartitioning is required based on Equation~\ref{eqn:end-to-end}. Two options are considered: (1) Variation of network speed between edge and cloud individually, and (2) Variation of CPU and memory stress on the edge individually. If either option requires DNN repartitioning, then it follows that the combination of options would also require repartitioning. 

Figure~\ref{fig:vgg-comm} shows the end-to-end latency for VGG-19 for all possible partitioning points (layer number shown on the x-axis) and the size of data is transferred between the edge and the cloud for each partitioning point when the network bandwidth is 20Mbps (typical upload speed in a broadband setting) and is dropped to 5Mbps (poorer quality upload speed); these values are chosen based on the literature~\cite{lockhart2020scission}. Other system resources, such as CPU and memory of the VM are fully available to the DNN. Each stacked bar represents the end-to-end latency of the DNN when partitioned at the layer identified on the x-axis. 
The blue and orange bars indicate the time taken for the partitions to execute on the edge and cloud, respectively. The time taken to transfer data is shown in grey. 
The partition point that has the lowest end-to-end latency is the optimal partition point (the shortest bar). 
Consider for example, the partitioning Layer 17 in Figure~\ref{fig:vgg-20}, which has the lowest end-to-end latency when the network speed between the edge and the cloud is 20Mbps. 
However, the optimal partitioning when the network speed drops to 5Mbps is Layer 22. 
Therefore, for VGG-19 it is noted that variation in network speed (a rise or drop) is a valid scenario for DNN repartitioning. 

Figure~\ref{fig:mobilenetv2-comm} shows similar results for MobileNetV2. Since this is a non-sequential model, individual layers are identified as a single number and blocks are represented as a range of layers (for example, layers 19-28 are a block). When the network speed is 20Mbps the optimal partitioning point is Layer 2 and when it drops to 5Mbps the partitioning point is Layer 35. 

It is thus observed that variation in network speed is a valid scenario for partitioning DNNs. 
Experiments (not presented due to space constraints) show that there was no effect from CPU and memory stress on the edge for DNN repartitioning.

\section{Reducing Edge Service Downtime}
\label{sec:technique}
This section presents the methods incorporated within \texttt{NEUKONFIG} that address the remaining two research questions posed in Section~\ref{sec:introduction} - (Q2) What is the edge service downtime that is associated with DNN repartitioning? To this end a baseline approach that makes use of a `Pause and Resume' technique from the literature when container-based applications need to be redeployed is presented. (Q3) How can edge service downtime be reduced when repartitioning a DNN? To this end, a `Dynamic Switching' approach is proposed.

Both the baseline and proposed approaches are considered in the context of video analytics where a camera captures video and then streams it to the edge server where it is pre-processed on the DNN partition. The output from the edge server is then transferred to the DNN partition on the cloud. The final result may be sent to the device. 
\subsection{Baseline Approach}

\begin{figure}[t]
    \centering
    \includegraphics[width=0.49\textwidth]{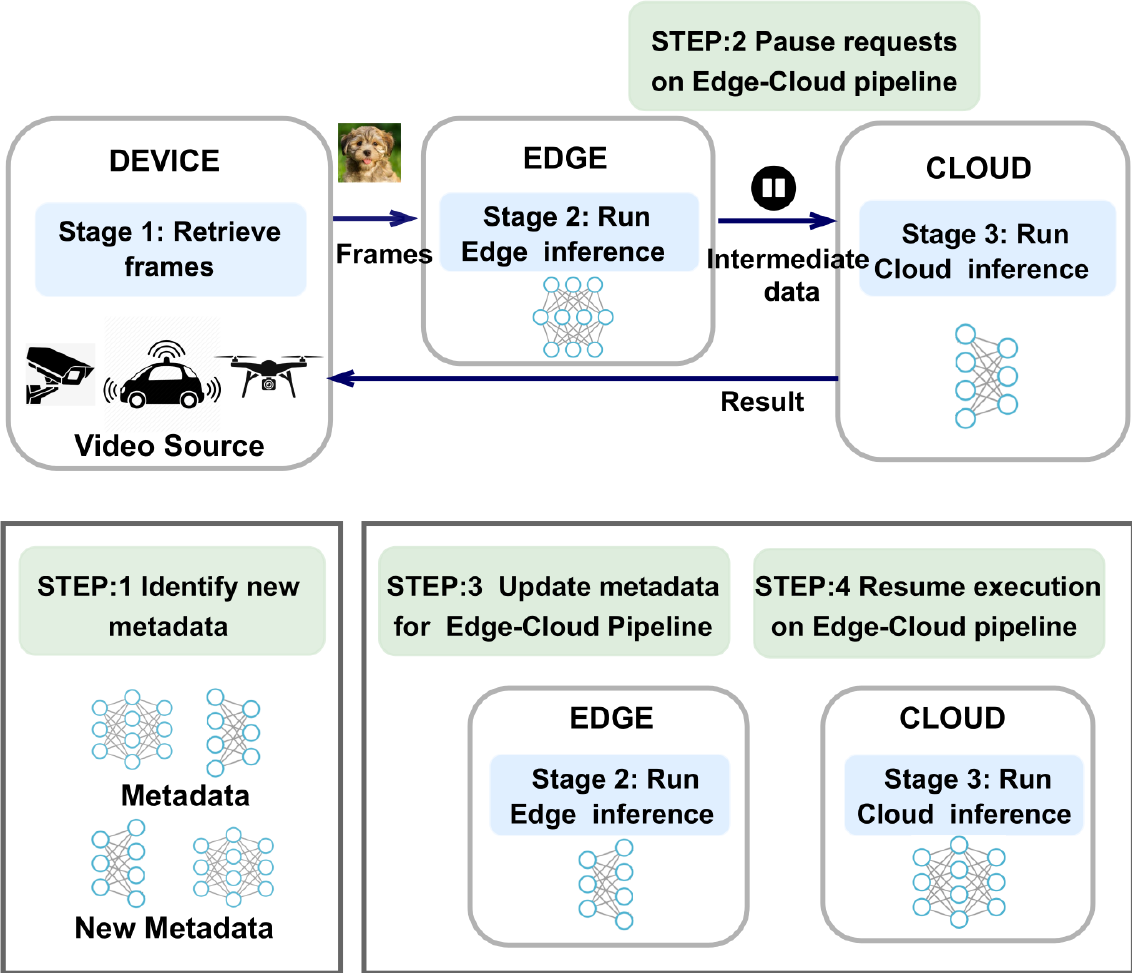}
    \caption{The baseline approach using the `Pause and Resume' technique for repartitioning DNNs}
    \label{fig:pause-approach}
\end{figure}

\begin{figure}[t]
    \centering
    \includegraphics[width=0.49\textwidth]{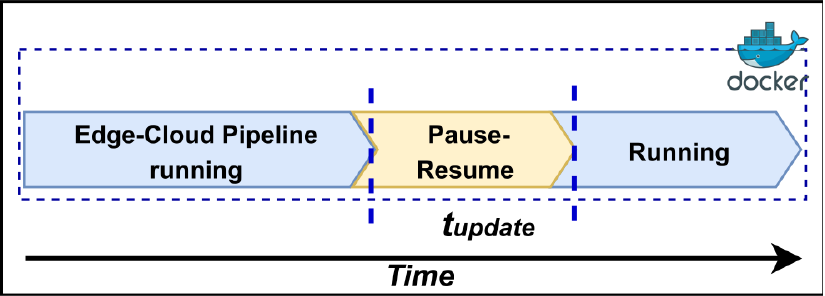}
    \caption{Sequence of activities for the baseline approach}
    \label{fig:sequence-pause}
\end{figure} 

The baseline approach relies on the `Pause and Resume' technique in which, when a change in network speed is observed, the running containers on the cloud and edge stop the execution of the DNN (pause). The container-based connection between the edge and the cloud will be referred to as the `edge-cloud pipeline' in this paper. The metadata (the new partitioning point for the DNN) is identified and provided to the DNN application. The application then resumes execution on the edge-cloud pipeline with the new metadata.   

The baseline approach consists of three sequential stages as shown in Figure~\ref{fig:pause-approach}: (i) Retrieve frames from a video stream generated by a video source, (ii) Run the edge-based DNN partition for inference and send the intermediate results to the cloud, and (iii) Run the cloud-based DNN partition for inference. The stages are connected using the ZeroMQ\footnote{https://zeromq.org/} asynchronous inter-process communication library that is employed to send and receive data.

The following four steps characterise the baseline approach as shown in Figure~\ref{fig:pause-approach}: 
(i) Identify new metadata, (ii) Pause processing requests on the edge-cloud pipeline, (iii) Update metadata for the edge-cloud pipeline, and (iv) \textit{Resume execution on the edge-cloud pipeline}, which are considered further.

(i) \textit{Identify new metadata}: The performance of the DNN may deteriorate as the network speed varies (whether independently or in combination with the system load on the edge). This requires the DNN to be repartitioned on new layers and distributed across the cloud and the edge. This step involves identifying the new metadata to maintain the required level of performance for the DNN based on the operational conditions of the edge-cloud system. To quickly adapt to run-time changes, optimal partition point may be identified using an estimation-based approach to predict the latency of individual layers of the DNN~\cite{wangcontext} or by using a real-time benchmarking approach~\cite{lockhart2020scission}. 

(ii) \textit{Pause processing requests on the edge-cloud pipeline}: The incoming requests for inferencing are temporarily paused between the edge and the cloud to update the DNN model on the edge and the cloud with new metadata. This is done so that the repartitioned DNN optimises its performance. 

(iii) \textit{Update metadata for the edge-cloud pipeline}: 
The new metadata is used to update the DNN model on the edge and cloud so that when the container can resume the execution of the DNN with the new partitioning configuration. 

(iv) \textit{Resume execution on the edge-cloud pipeline}: The execution of the paused container is resumed with the DNN model executing based on the new metadata. Requests for inferencing input images for example can be resumed on the edge-cloud pipeline. 

When the container is paused no incoming requests from the devices can be processed on the edge. The downtime on the edge server $t_{downtime}$ is $t_{update}$ (as shown in Figure~\ref{fig:sequence-pause}, which is the time the container on the edge and cloud is paused until the containers can resume execution with the new DNN partitions:
\begin{equation}
t_{downtime}=t_{update}
\label{eqn:pause-n-resume}
\end{equation}

This approach may be advantageous in environments where the edge has limited resource availability and high availability is not necessarily a priority. More sophisticated approaches that reduce downtime may require a larger amount of resources (memory and CPU availability).

\subsection{Dynamic Switching Approach}
In this section, the motivation for the design of the proposed Dynamic Switching approach, the anatomy of the approach and the scenarios relevant to the approach are presented. 

\subsubsection{Motivation}
The following two observations in relation to DNN repartitioning motivate the design of the proposed approach that aims to reduce edge server downtime:

(i) \textit{Redeployment of repartitioned DNNs across the edge and the cloud must be rapid}: This is because the edge is dynamic (network conditions may change rapidly over time and device mobility affects which edge server should service a device).  Frequent repartitioning would be required in such a transient environment and would require the software deployed in these environments to rapidly adapt to changes in the operational conditions.

(ii) \textit{Redeployment approaches must be proactive}: Current approaches presented in the literature that address DNN partitioning and deployment are based on historic and profiled data that is available to the system~\cite{skarlat2017towards, yousefpour2019fogplan, murtaza2020qos}. However, such approaches are not suitable in the context of repartitioning DNNs when a running DNN application needs to cope with current system changes to maintain performance.

Therefore, the approach aims to be more rapid than the baseline approach to cope with operational changes in the cloud-edge environment and proactive and practical to reduce the impact of edge service downtime. 

\begin{figure}[t]
    \centering
    \includegraphics[width=0.49\textwidth]{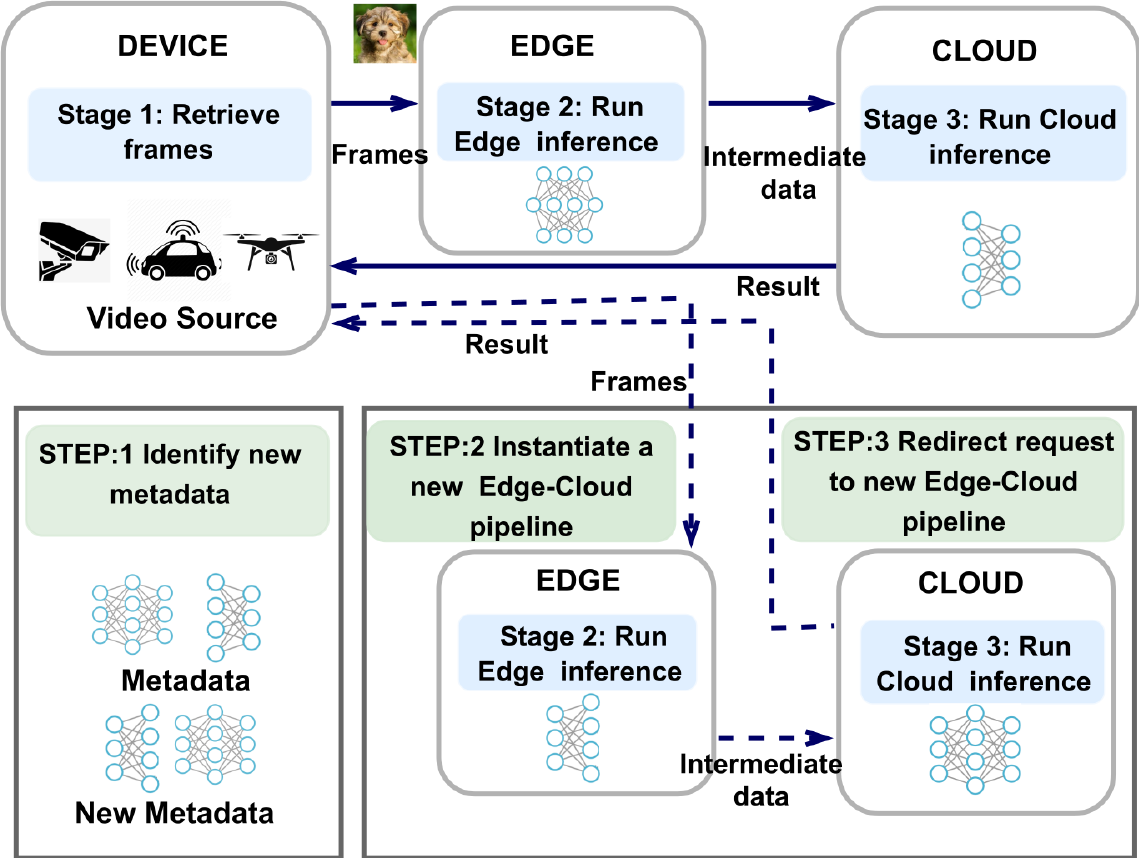}
    \caption{Dynamic Switching approach for reducing edge server downtime}
    \label{fig:switch-approach}
\end{figure} 

\subsubsection{Approach}

The approach has three sequential stages as discussed for the baseline approach. 
The Dynamic Switching approach consists of three steps as shown in Figure~\ref{fig:switch-approach}: (i) Identify new metadata (ii) Instantiate a new or use a different edge-cloud pipeline, and (iii) Redirect requests to the new or different edge-cloud pipeline:

(i) \textit{Identify new metadata}: Identifying the new metadata is based on the baseline approach discussed previously.

(ii) \textit{Instantiate a new edge-cloud pipeline}: There are multiple scenarios in which a repartitioned DNN can be executed. They are either by initialising new containers on the edge and cloud with the new DNN partitions or by running the DNN partitions on an already existing (redundant) pipeline. These cases will be considered later. 
(iii) \textit{Redirect requests to the new edge-cloud pipeline}: Device requests (for example, inference of video frames) are redirected or switched to the new edge-cloud pipeline. The background pipeline may be terminated depending on the scenario considered next. 

One difference between the baseline approach using Pause and Resume and the Dynamic Switching approach is that for the former during the downtime no device requests can be processed; the edge server is completely interrupted. However, in the proposed approach, the edge server downtime refers to an interval of time when the quality of service on the edge is degraded, but is still operational. 

\subsubsection{Scenarios}
Two scenarios (Scenario A and Scenario B) are considered in the Dynamic Switching approach. These are based on when a different edge-cloud pipeline is initialised. In
each scenario, there are two cases depending on whether the second pipeline runs in a
new container (Case 1) or the existing container (Case 2). Case 1 will require spare resources to be available on the edge server so that a new container can simultaneously run with the container of the first pipeline.

\textit{Scenario A}: Here, a redundant edge-cloud pipeline is always running for the application. One pipeline will be used for processing the device requests for a given DNN partition on the edge and the cloud. The second pipeline remains idle. When there is a change in the network speed, the device requests are redirected to the redundant pipeline, thus reducing the downtime as observed in the baseline approach. The downtime is: 

\begin{equation}
 t_{downtime} = t_{switch}
\label{eqn:switch-a}
\end{equation}

\begin{figure}[t]
    \centering
    \includegraphics[width=0.49\textwidth]{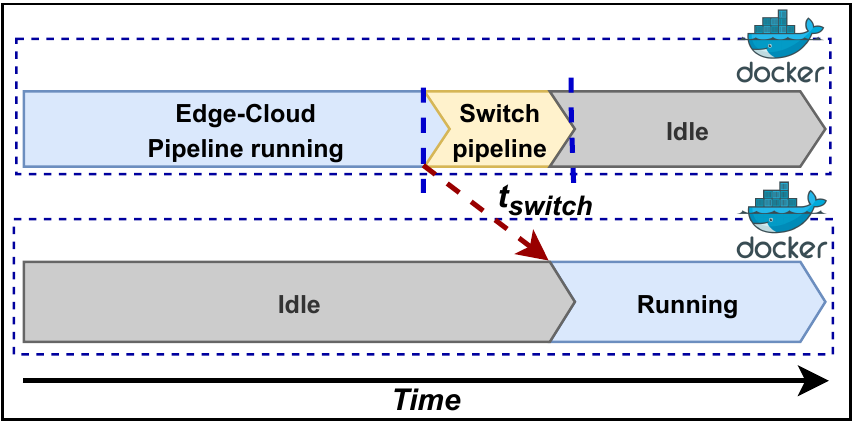}
    \caption{Sequence of activities for Scenario A, Case 1 in the Dynamic Switching approach}
    \label{fig:sequence-scenario-a}
\end{figure} 

\begin{figure}[t]
    \centering
    \includegraphics[width=0.49\textwidth]{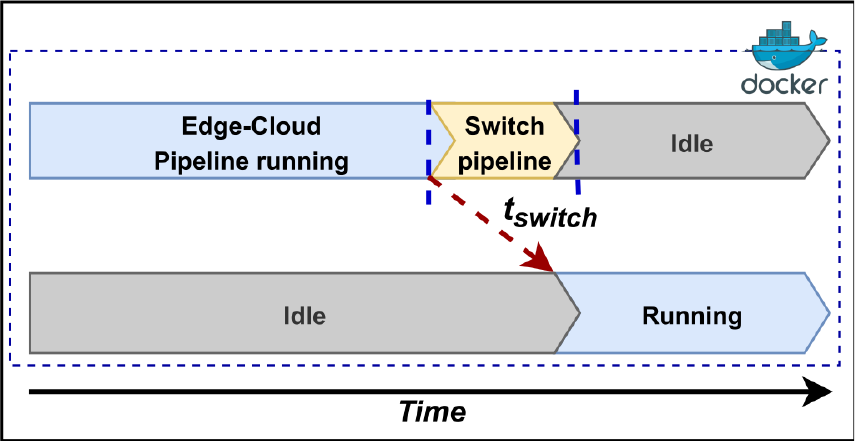}
    \caption{Sequence of activities for Scenario A, Case 2 in the Dynamic Switching approach}
    \label{fig:sequence-scenario-b}
\end{figure} 

\begin{figure}[t]
    \centering
    \includegraphics[width=0.49\textwidth]{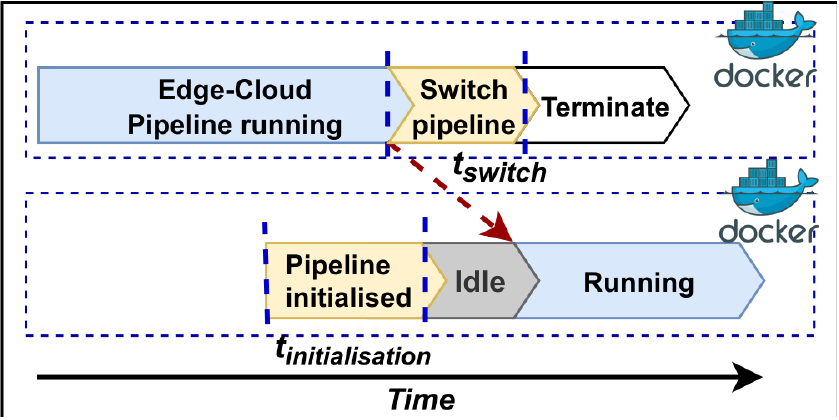}
    \caption{Sequence of activities for Scenario B, Case 1 in the Dynamic Switching approach}
    \label{fig:sequence-scenario-b-1}
\end{figure}

\textit{Scenario B}: In Scenario B, a new edge-cloud pipeline is initialised only when a change in network speed is observed. Unlike Scenario A, a redundant pipeline does not always exist and it is hypothesised that Scenario B will require fewer resources and thus will be more resource efficient. 
In Case 1, when there is a change in network speed, new containers will be initialised (building and setting up application containers) on both the edge and the cloud and then requests are redirected to the new edge-cloud pipeline.
The time for initialisation is denoted as $t_{initialisation}$ and the time to switch is denoted as $t_{switch}$ as shown in Figure~\ref{fig:sequence-scenario-b-1}. The downtime in this case is:
\begin{equation}
t_{downtime}=t_{initialisation} +t_{switch}
\label{eqn:switch-scenario-a}
\end{equation}
The above scenario is designed to reduce downtime of edge service, but this approach will require spare resources to be available on the edge server so that a new container can simultaneously run with the container of the first pipeline.

\begin{figure}[t]
    \centering
    \includegraphics[width=0.49\textwidth]{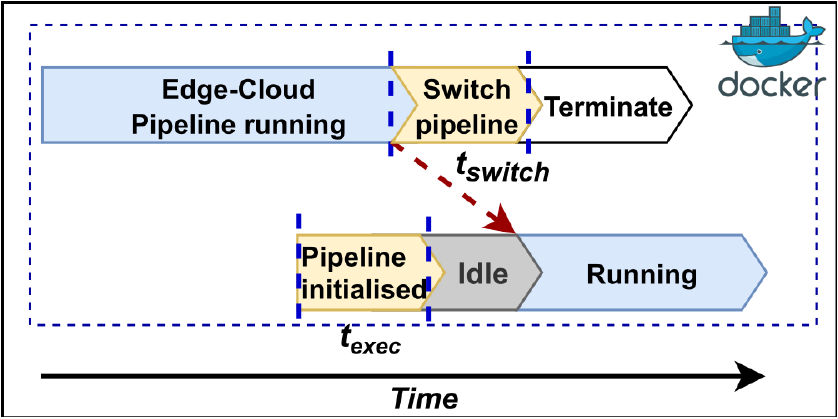}
    \caption{Sequence of activities for Scenario B, Case 2 in the Dynamic Switching approach}
    \label{fig:sequence-scenario-b-2}
\end{figure}

In Case 2, a new pipeline is initialised within the existing containers taking
time $t_{exec}$ as shown in Figure~\ref{fig:sequence-scenario-b-2}.
This case eliminates the time required to build and initialise containers for the second pipeline and at the same time requires fewer resources on the edge. The downtime in this case is:
\begin{equation}
t_{downtime}=t_{exec} +t_{switch}
\label{eqn:switch-scenario-b}
\end{equation}

\section{Experimental Study}
\label{sec:experiments}
This section presents the setup used by \texttt{NEUKONFIG} for gathering empirical data from an experimental edge-cloud testbed to evaluate the proposed Dynamic Switching approach for reducing edge server downtime against the baseline approach. 

\subsection{Experimental Setup}
The setup consists of: (i) A device modelled as a video camera with on-device capabilities of a Raspberry Pi 3B+ with a 8 core ARM processor and 1 GB RAM, (ii) An edge server with a 4 core x86\_64 processor and 8 GB RAM running Ubuntu 18 LTS, and (iii) A cloud server with a 8 core x86\_64 processor and 16 GB RAM running Ubuntu 18 LTS. Docker 18.09-ce is installed on the device, edge and cloud servers. 

For experimentation, the CPU and memory availability are controlled on the edge server using stress-ng. The outbound network traffic between the client, edge and cloud server are emulated using Linux Traffic Control (\texttt{tc}). 
To emulate real network conditions between the edge and cloud as seen in fibre broadband, an average network speed of 20Mbps and a network latency of 20ms are used\footnote{https://www.ofcom.org.uk}.
A WiFi or Ethernet connection between the device and the edge server is assumed.

\subsection{Results}
The aim of the experiments carried out is to answer questions Q2 and Q3 posed above. 
Q1 was addressed in Section~\ref{sec:background}.
To address the above questions the experiments will highlight the service downtime for the baseline (Pause and Resume) and the Dynamic Switching approaches for different scenarios considered in Section~\ref{sec:technique}.

\begin{figure}[t]
\begin{center}
	\subfloat[When network speed changes to 20Mbps]
	{\label{fig:downtime-pause5Mb}
	\includegraphics[width=0.237\textwidth]
	{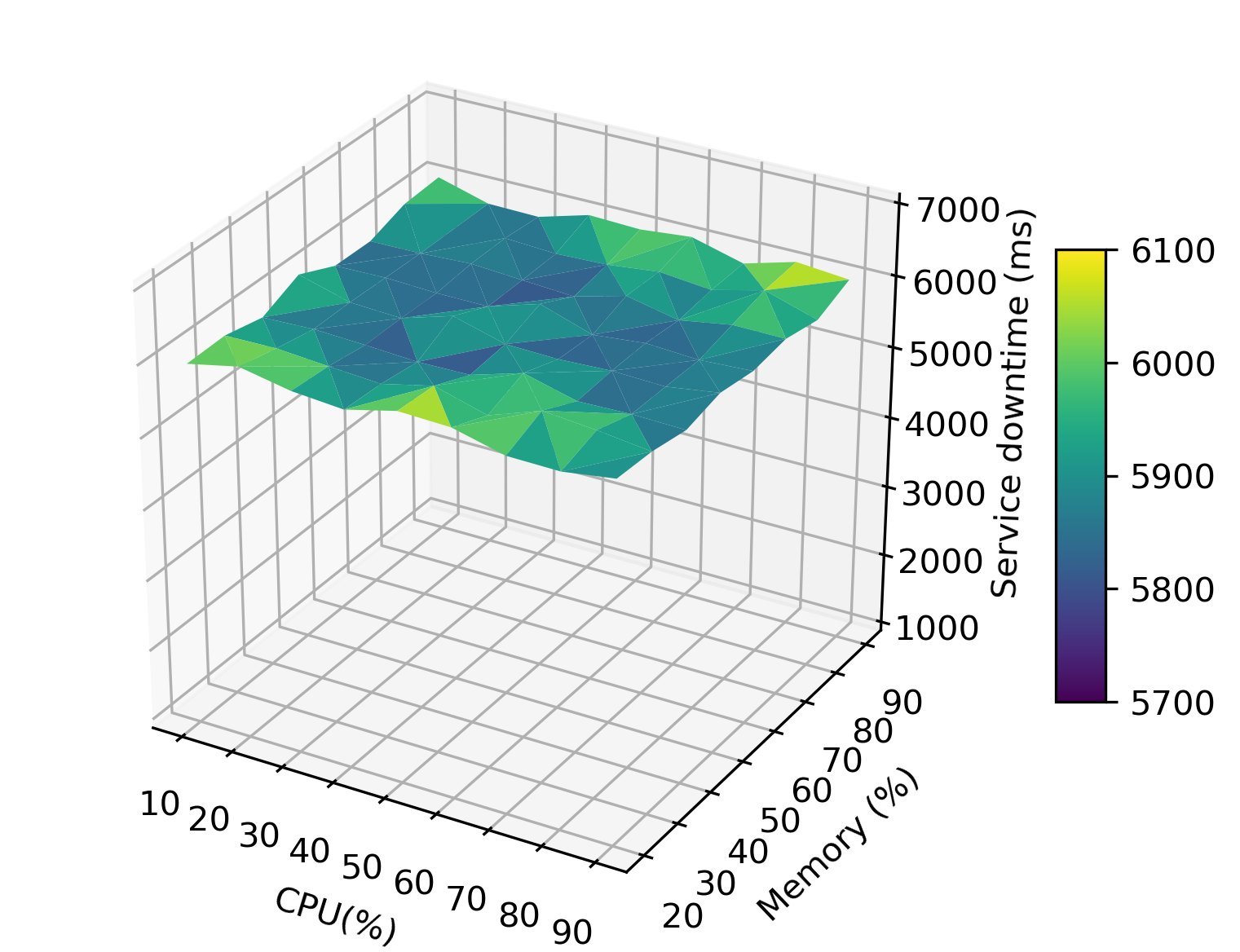}}
	\hfill
	\subfloat[When network speed changes to 5Mbps]
	{\label{fig:downtime-pause20Mb}
	\includegraphics[width=0.237\textwidth]
	{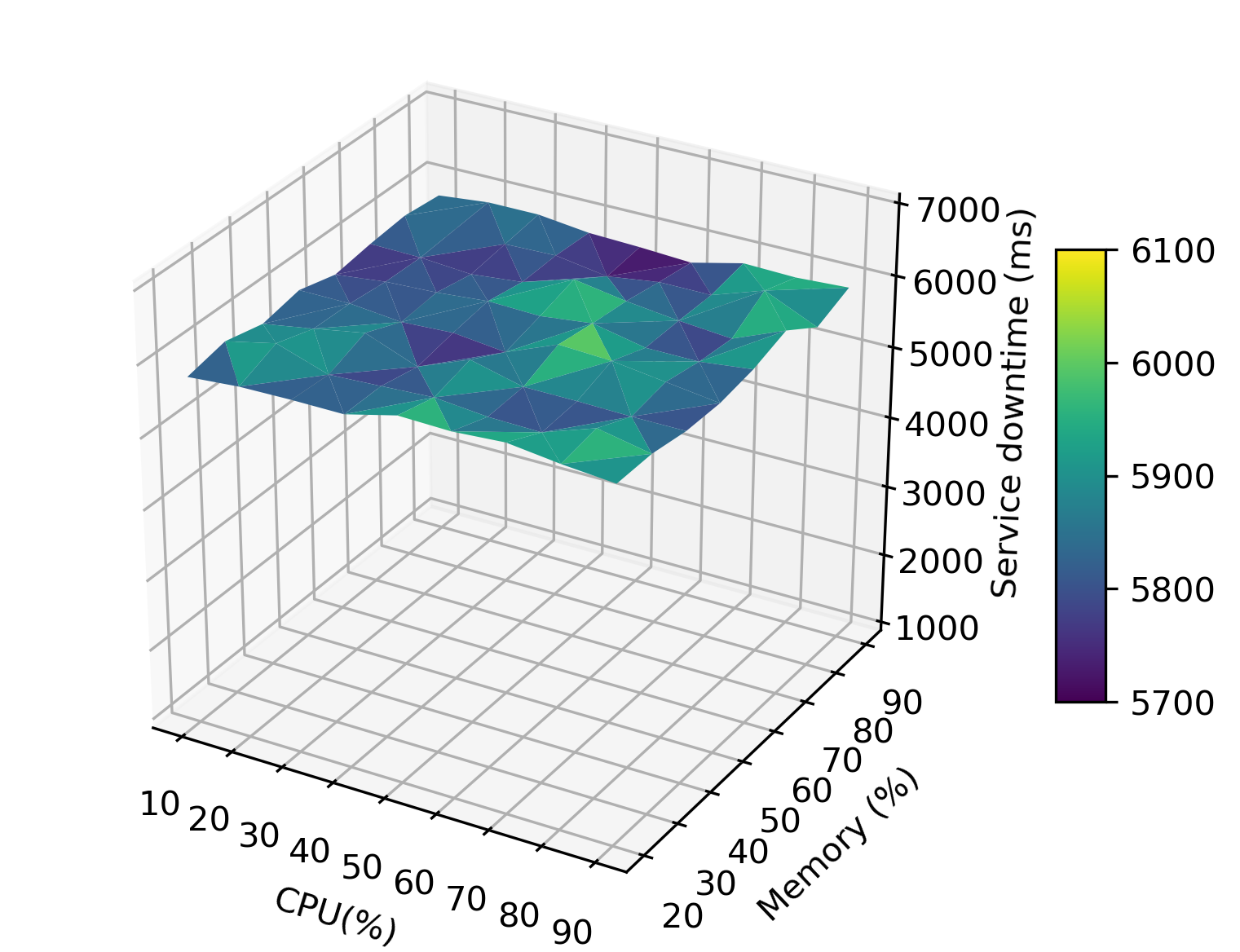}}
\end{center}
\caption{Edge service downtime for Pause and Resume approach when network speed changes.}
\label{fig:downtime-pause}
\end{figure}

Service downtime observed at the edge while repartitioning the DNNs using the baseline approach is captured in Equation~\ref{eqn:pause-n-resume}. The downtime is incurred since the DNN partition on the edge and cloud servers need to be updated. Figure~\ref{fig:downtime-pause} shows the service downtime for the baseline when network speed changes (Figure~\ref{fig:downtime-pause20Mb} is for a change from 5Mbps to 20Mbps and Figure~\ref{fig:downtime-pause5Mb} is for 20Mbps to 5Mbps). The x-axis of the graphs denotes percentage of CPU availability, y-axis denotes percentage of memory available on Edge server, and z-axis denotes service downtime in milliseconds. A downtime of around 6 seconds is observed when using the baseline approach when the network speed changes. This is a relatively large period of time for an edge server to be disrupted. During this time the edge can offer no service to the client; no frames sent from the device to the
edge will be processed. 

The downtime is incurred since inference needs to be paused on both the edge and the cloud while updating the partitions. CPU and memory availability and varying network conditions do not change the service downtime. 

When the available memory was less than or equal to 10\%, the DNN partitions could not be executed on the edge and so no results are shown for 10\% memory availability.

The downtime in two scenarios, referred to as Scenario A (when a second edge-cloud pipeline is always running) and Scenario B (when a second edge-cloud pipeline is created only when needed) as presented in Section~\ref{sec:technique} are considered for the Dynamic Switching approach. 

\begin{figure}[t]
\begin{center}
	\subfloat[When network speed changes to 20Mbps]
	{\label{fig:downtime-20Mb-a}
	\includegraphics[width=0.237\textwidth]
	{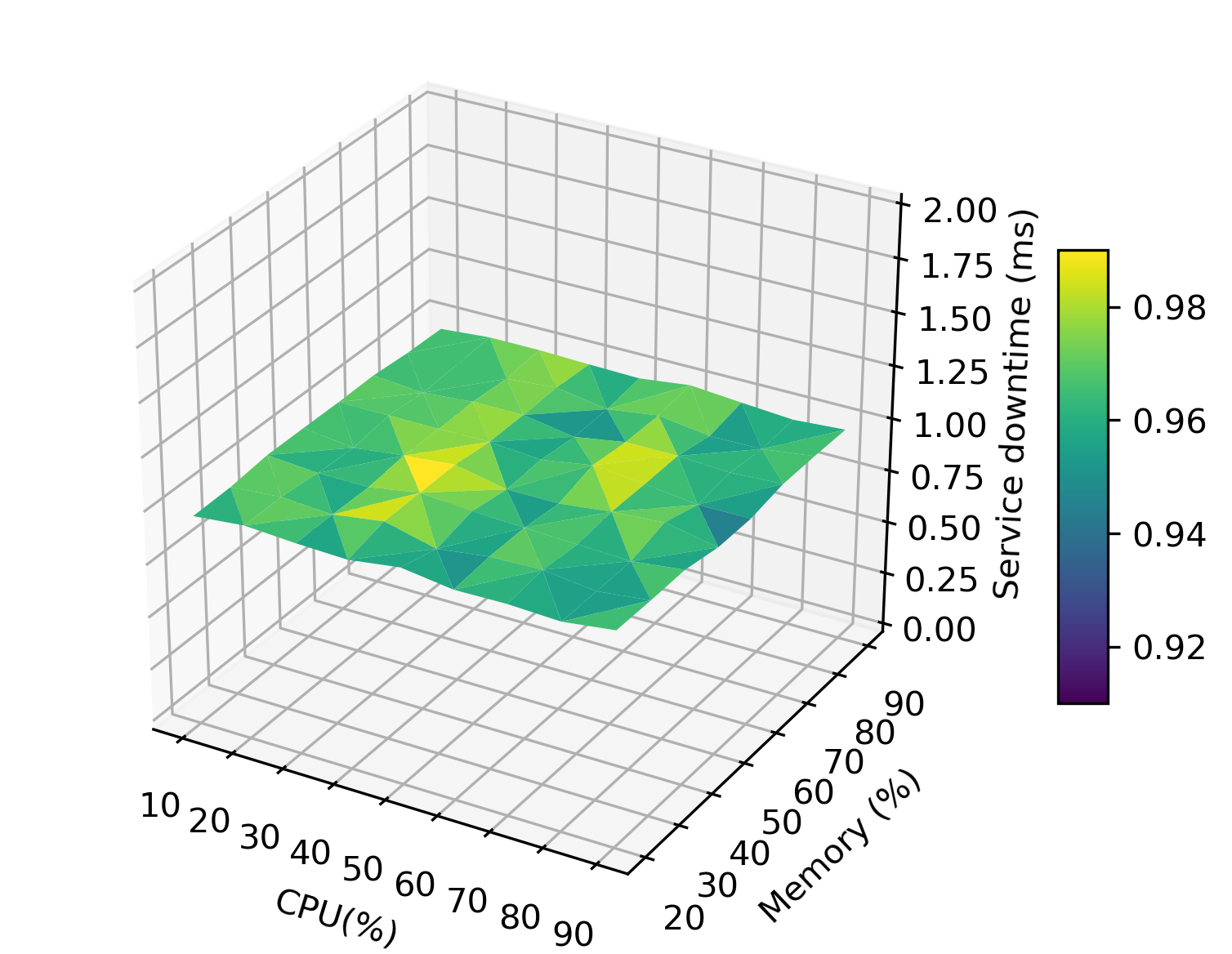}}
	\hfill
	\subfloat[When network speed changes to 5Mbps]
	{\label{fig:downtime-5Mb-a}
	\includegraphics[width=0.237\textwidth]
	{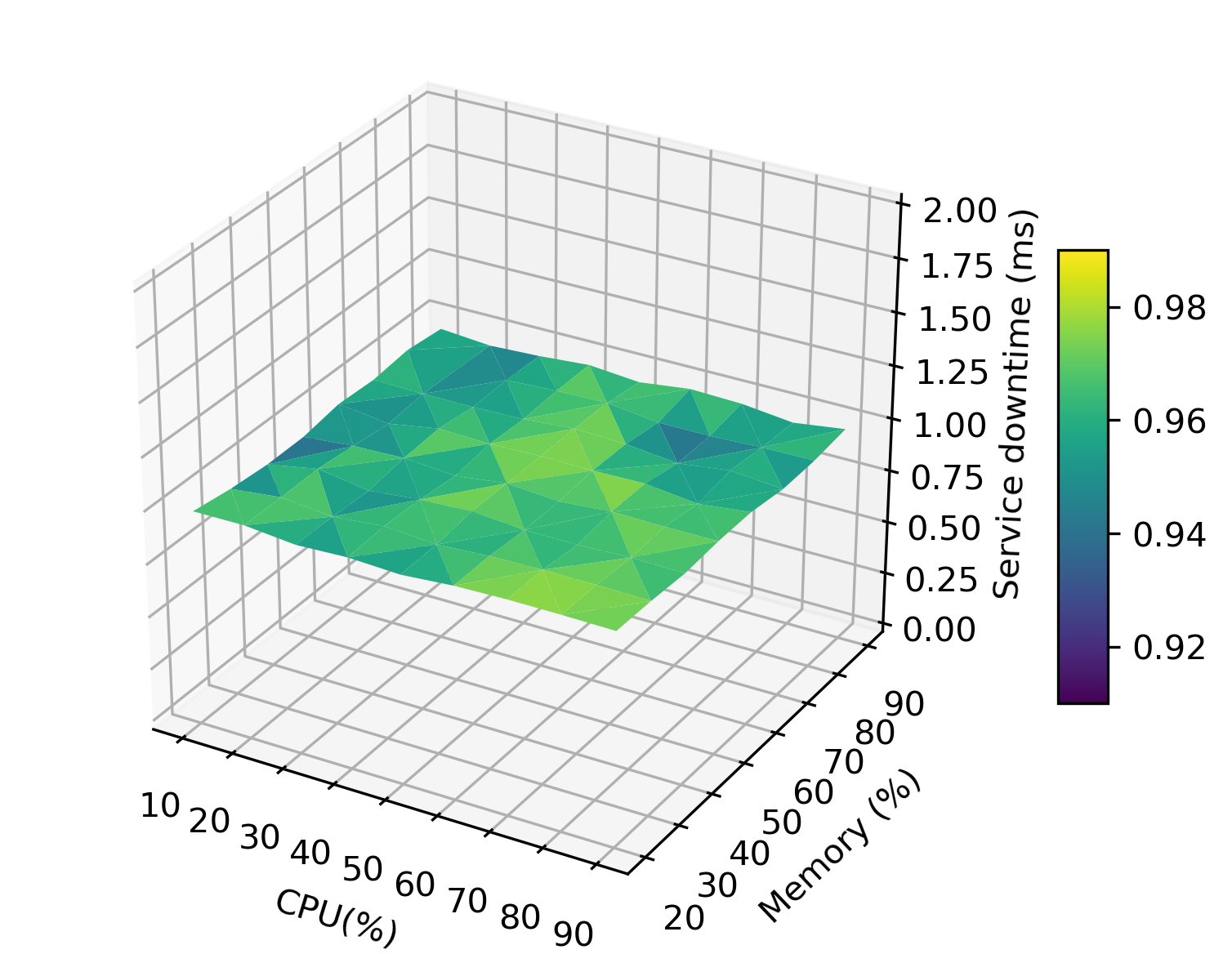}}
\end{center}
\caption{Edge service downtime for Dynamic Switch approach when network speed changes in Scenario A.}
\label{fig:downtime-scenario-a}
\end{figure}

 Figure~\ref{fig:downtime-scenario-a} shows service downtime for Scenario A, which involves switching incoming requests of a device to an existing pipeline.
 Case 1 and Case 2 have the same downtime because container initialisation is
 already complete (based on Equation~\ref{eqn:switch-a}). Figure~\ref{fig:downtime-20Mb-a} and Figure~\ref{fig:downtime-5Mb-a} show the edge service downtime when network speed changes to 20Mbps and 5Mbps, respectively. A relatively low service down time of less than 0.98ms under different CPU and memory availabilities are noted.

\begin{figure*}[t]
\begin{center}
	\subfloat[Case 1: network speed is 20Mbps]
	{\label{fig:downtime-20Mb-b-1}
	\includegraphics[width=0.237\textwidth]
	{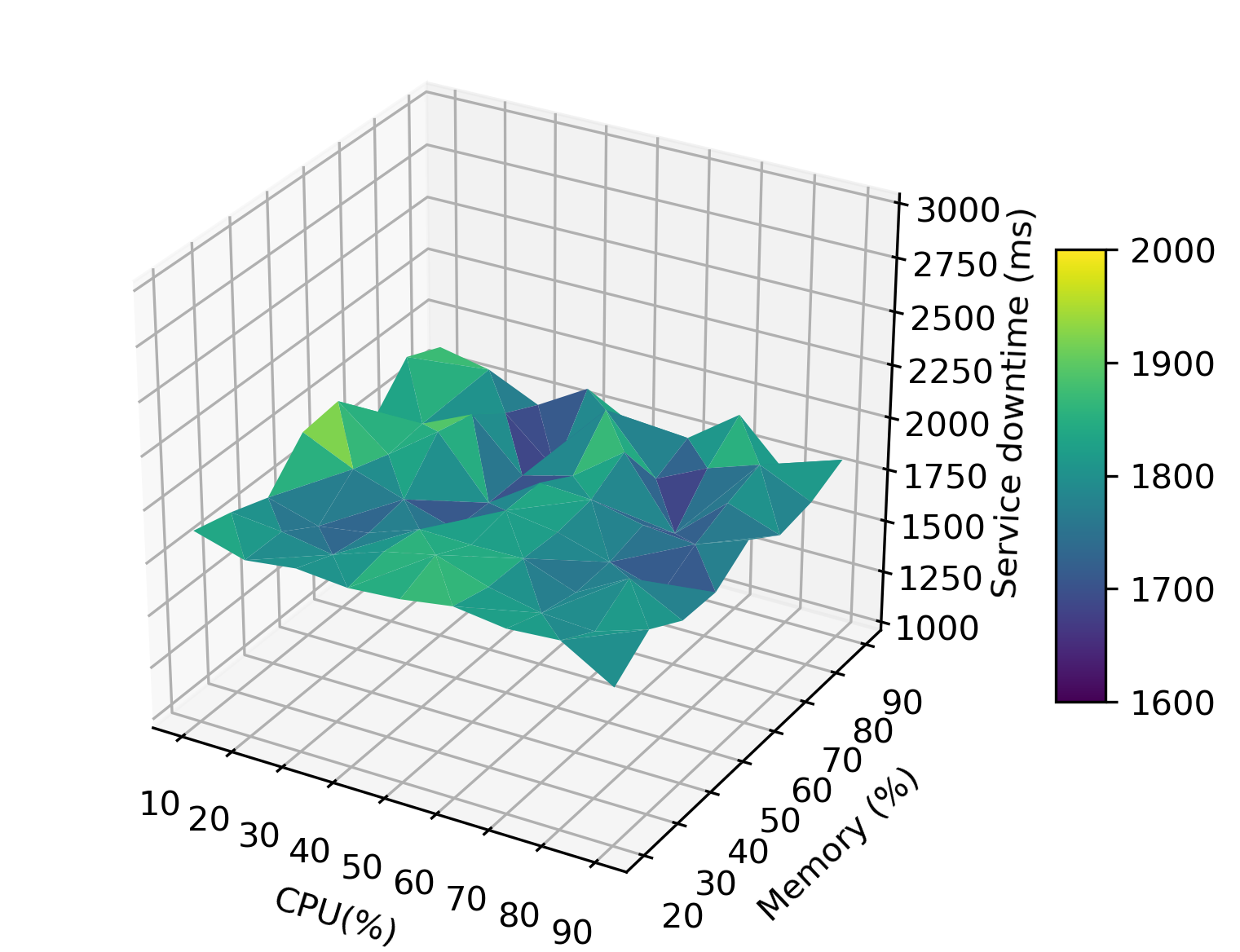}}
	\hfill
	\subfloat[Case 1: network speed is 5Mbps]
	{\label{fig:downtime-5Mb-b-1}
	\includegraphics[width=0.237\textwidth]
	{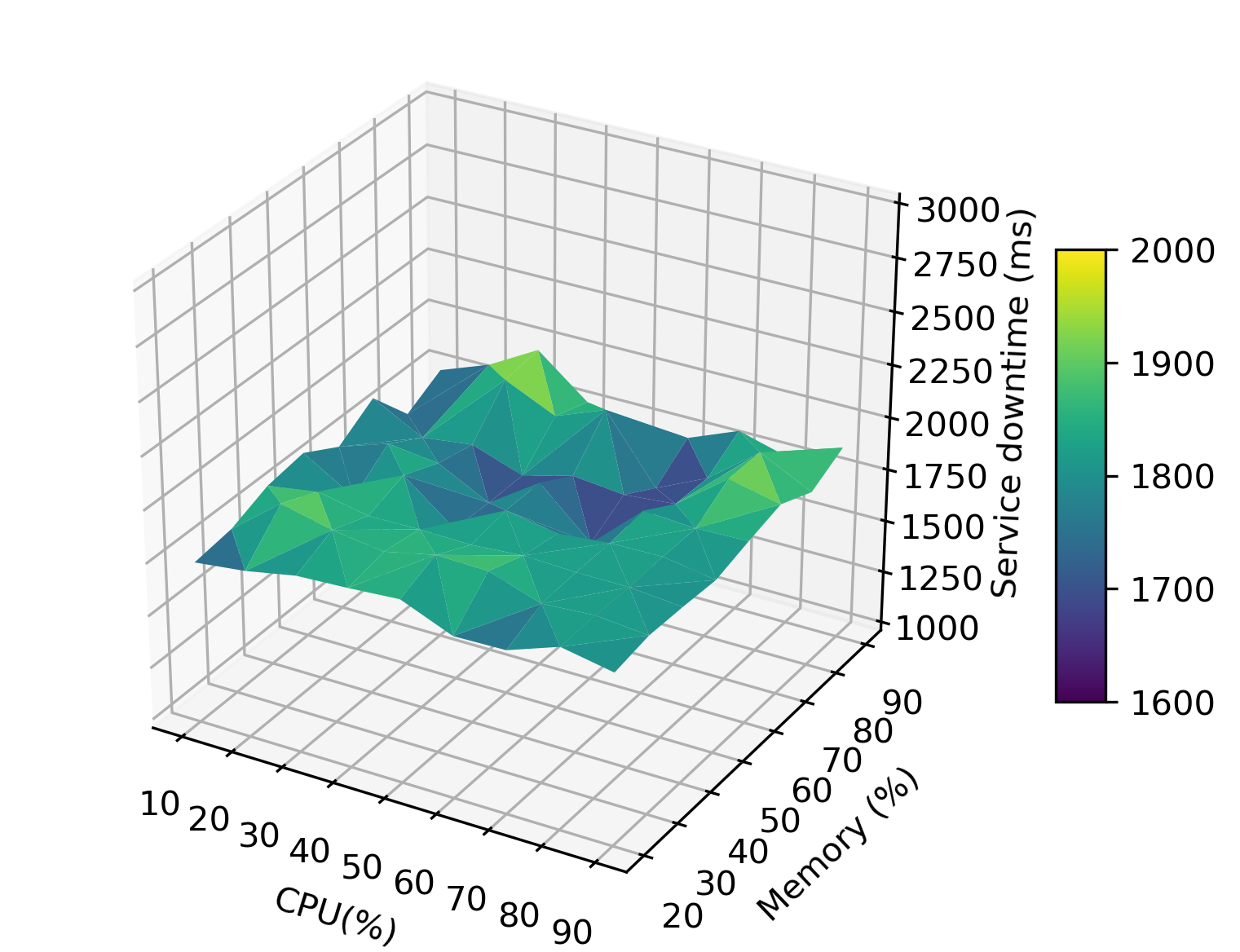}}
	\hfill
	\subfloat[Case 2: network speed is 20Mbps]
	{\label{fig:downtime-20Mb-b-2}
	\includegraphics[width=0.237\textwidth]
	{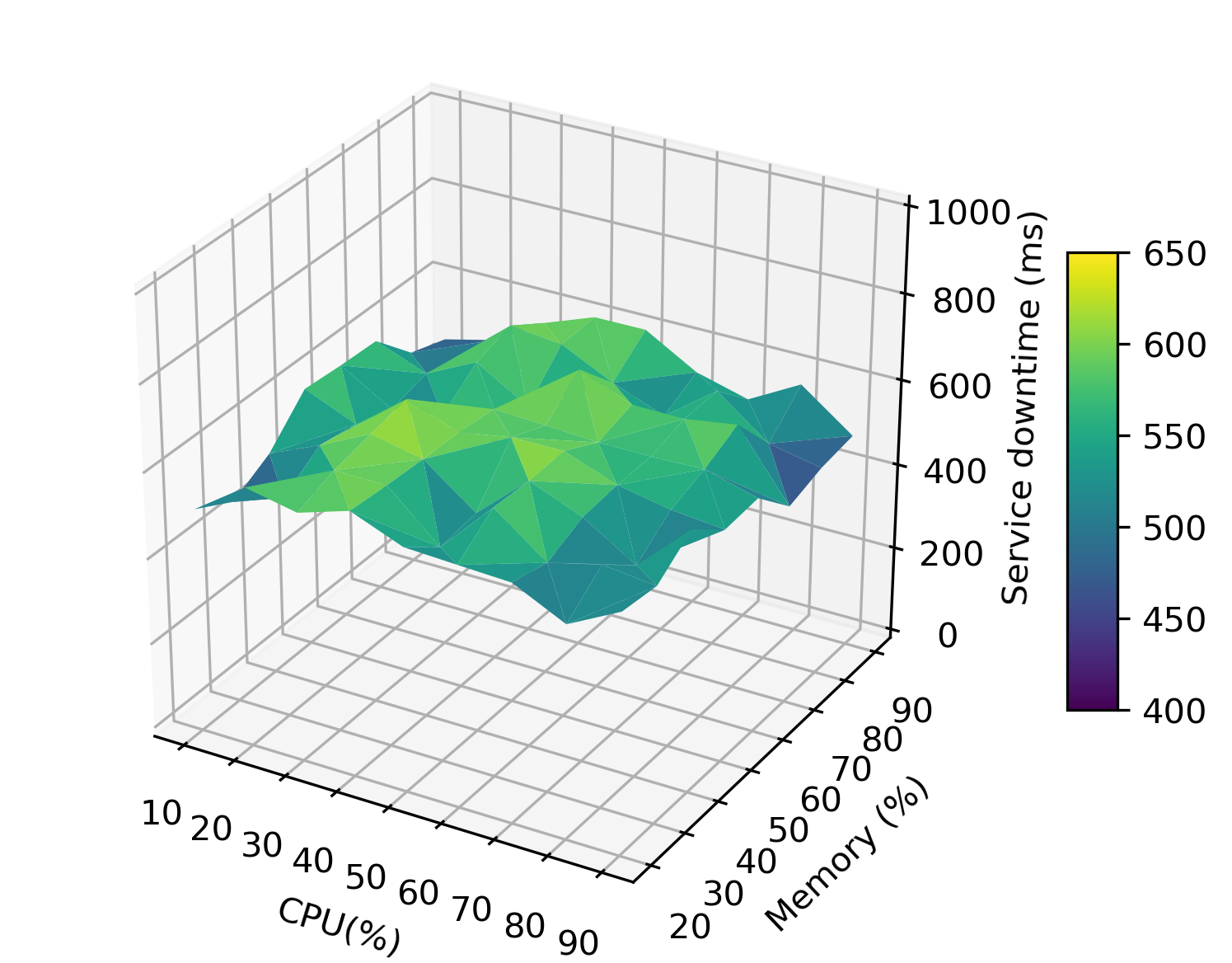}}
	\hfill
	\subfloat[Case 2: network speed is 5Mbps]
	{\label{fig:downtime-5Mb-b-2}
	\includegraphics[width=0.237\textwidth]
	{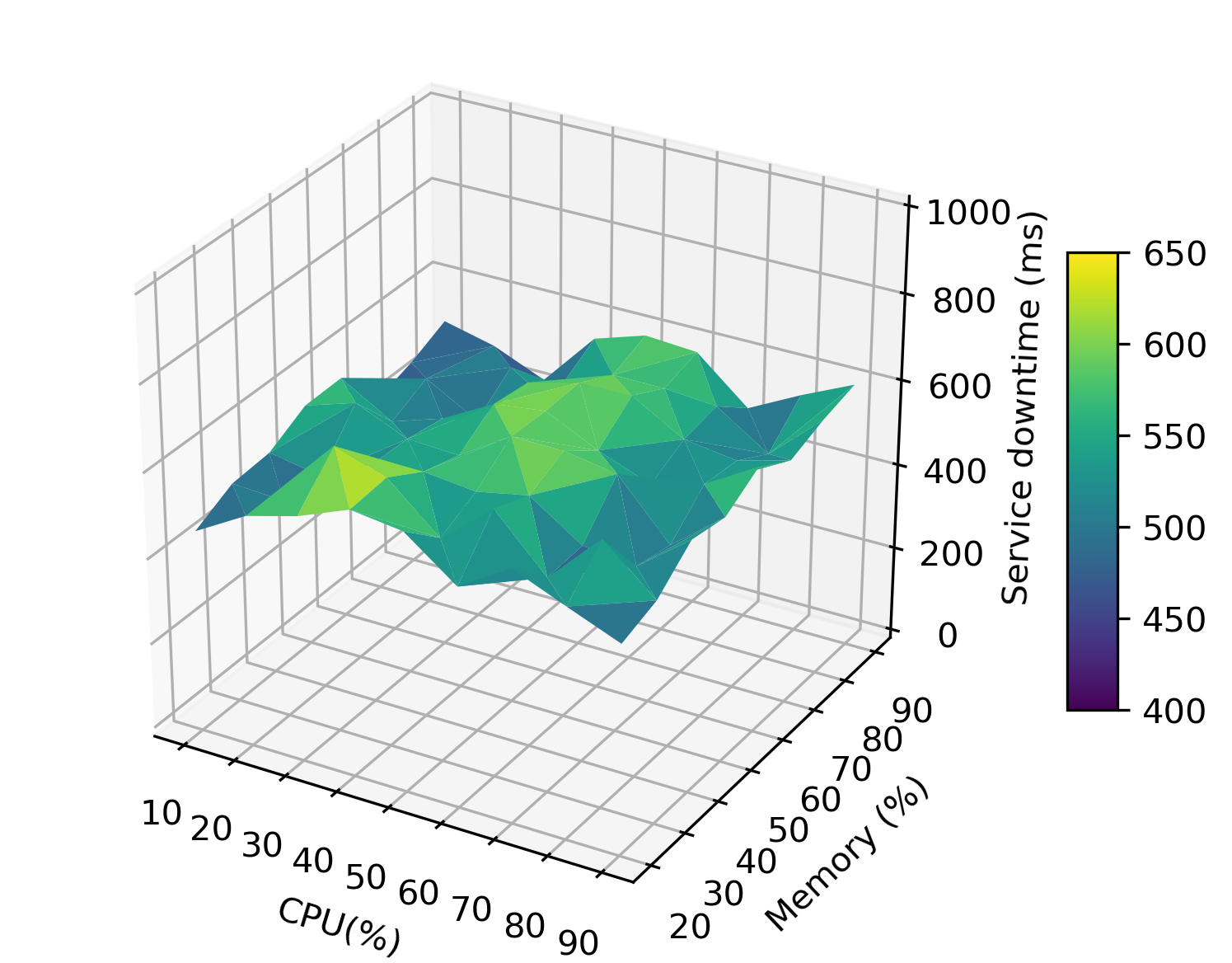}}
\end{center}
\caption{Edge service downtime for Case 1 and Case 2 of the Dynamic Switching approach in Scenario B.}
\label{fig:downtime-scenario-b}
\end{figure*}

Figure~\ref{fig:downtime-scenario-b} shows service downtime for Scenario B by taking two cases, Case 1 (new pipeline is created on a new container) and Case 2 (new pipeline is created on the same container) into account. Figure~\ref{fig:downtime-20Mb-b-1} and Figure~\ref{fig:downtime-5Mb-b-1} show service down time of nearly 1.9 seconds in Case 1. This is because a new (Docker) container needs to be built and initialised. 

The container image was optimised to reduce the time to set up a new edge-cloud pipeline for the Dynamic Switching approach and to reduce the memory footprint on the edge server. 
All libraries that are required to run a pipeline, such as TensorFlow and Pyzmq, are pre-installed in a base image and stored in a local cache on the edge and cloud servers. Only the DNN application specific resources are initialised in the new pipeline and the base image is shared between the pipelines. Thus, the container image size was reduced to 575 MB making it more compact for a faster build and startup when initialising the new pipeline.

Similarly, Figure~\ref{fig:downtime-20Mb-b-2} and Figure~\ref{fig:downtime-5Mb-b-2} show a service down time of 0.6 second when network speed changes in Case 2. The downtime is lower than Case 1 since a new pipeline is initialised within the same running container as the first pipeline, thereby eliminating the need to build and run a new container. 

In the baseline Pause and Resume approach, downtime implies that no frames can be processed on the edge. However, in Dynamic Switching, since the initial edge-cloud pipeline continues to run, it will process frames at sub-optimal performance. Therefore, we observe the frame drop rate at the edge during the downtime incurred (0.98ms for Scenario A; 1.9 seconds for Scenario B, Case 1; 0.6 second for Scenario B, Case 2). 

\begin{figure*}[t]
\begin{center}
	\subfloat[Incoming frame rate set to 5FPS]
	{\label{fig:fps-5-20mb}
	\includegraphics[width=0.237\textwidth]
	{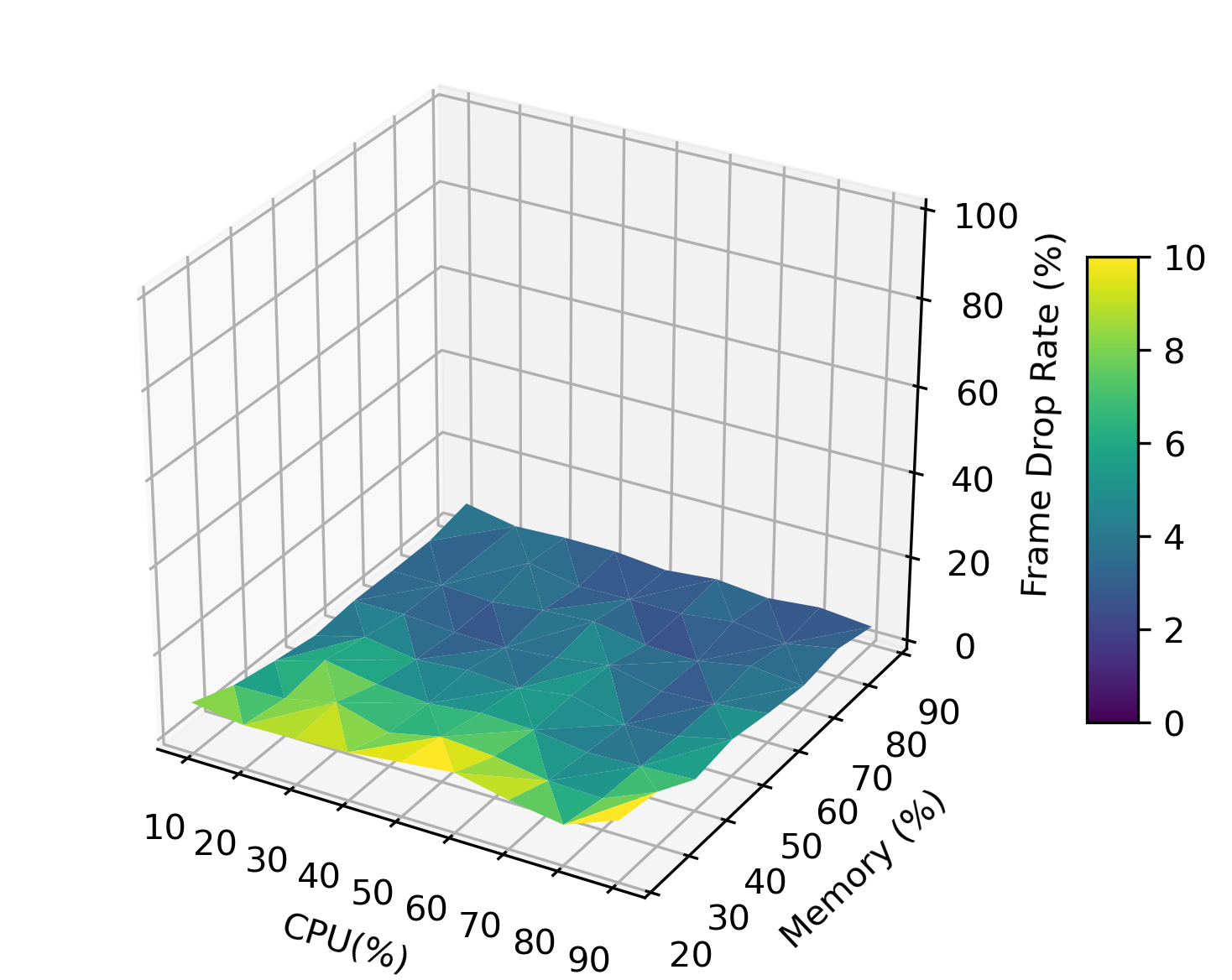}}
	\hfill
	\subfloat[Incoming frame rate set to 10FPS]
	{\label{fig:fps-10-20mb}
	\includegraphics[width=0.237\textwidth]
	{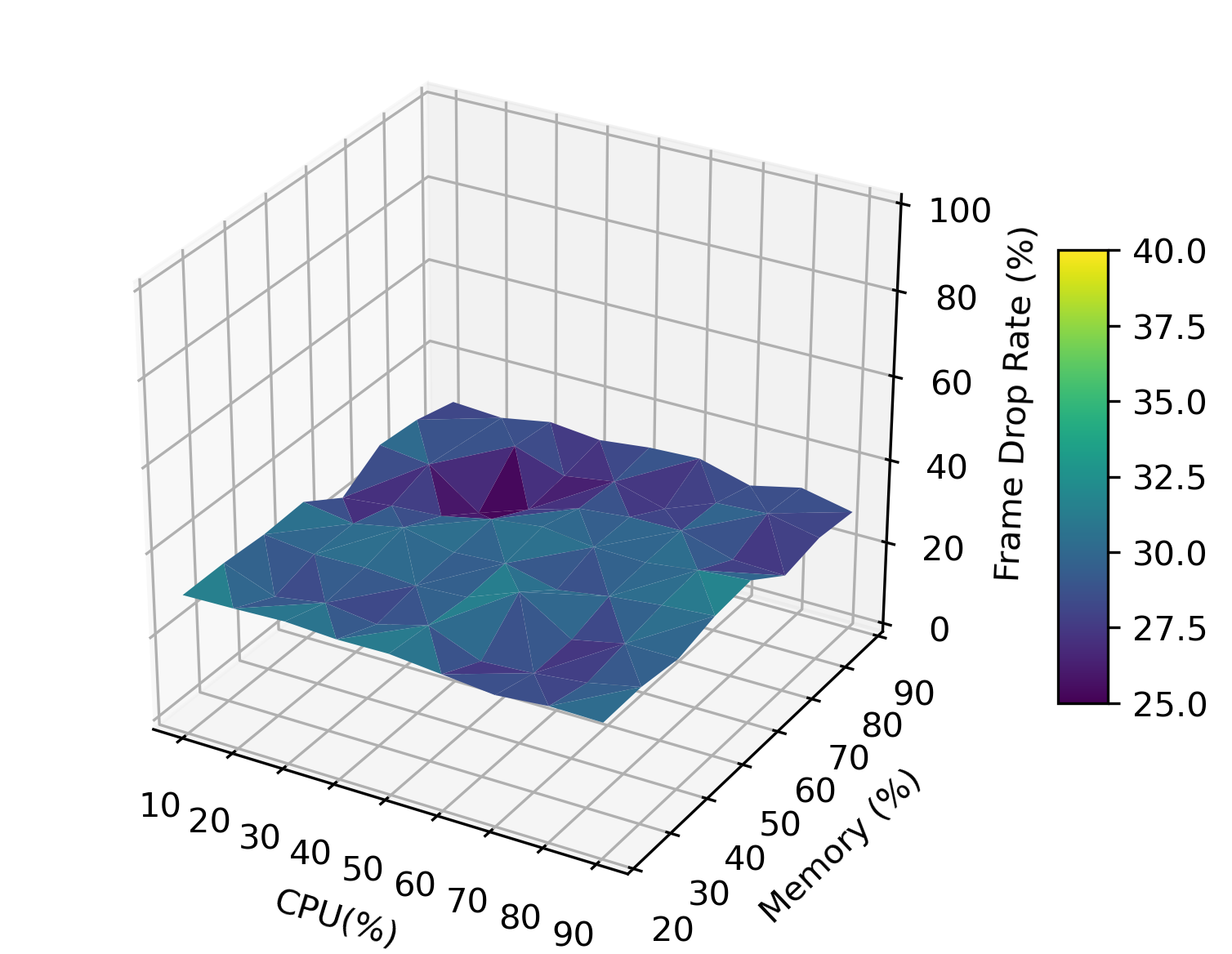}}
	\hfill
	\subfloat[Incoming frame rate set to 15FPS]
	{\label{fig:fps-15-20mb}
	\includegraphics[width=0.237\textwidth]
	{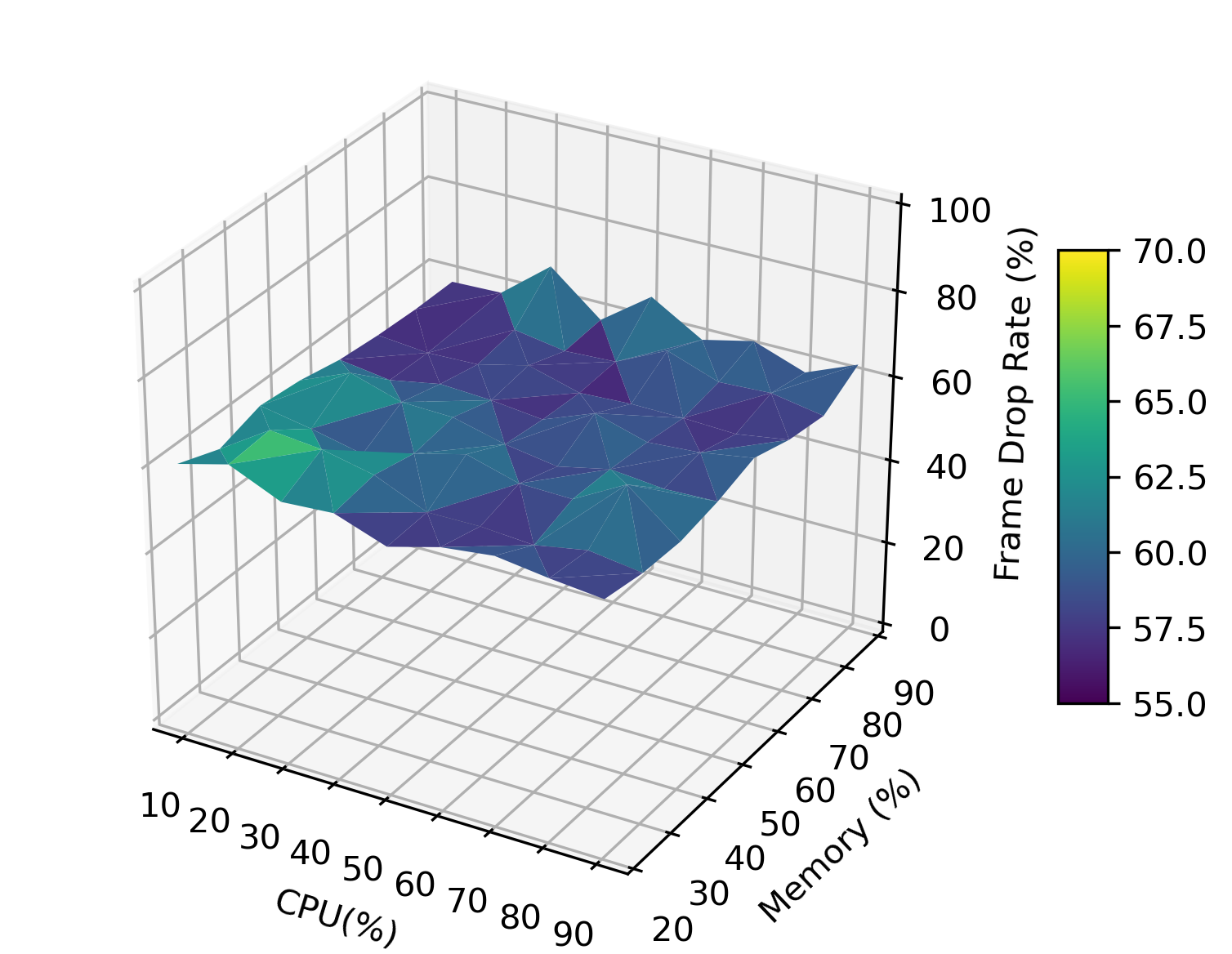}}
	\hfill
	\subfloat[Incoming frame rate set to 20FPS]
	{\label{fig:fps-20-20mb}
	\includegraphics[width=0.237\textwidth]
	{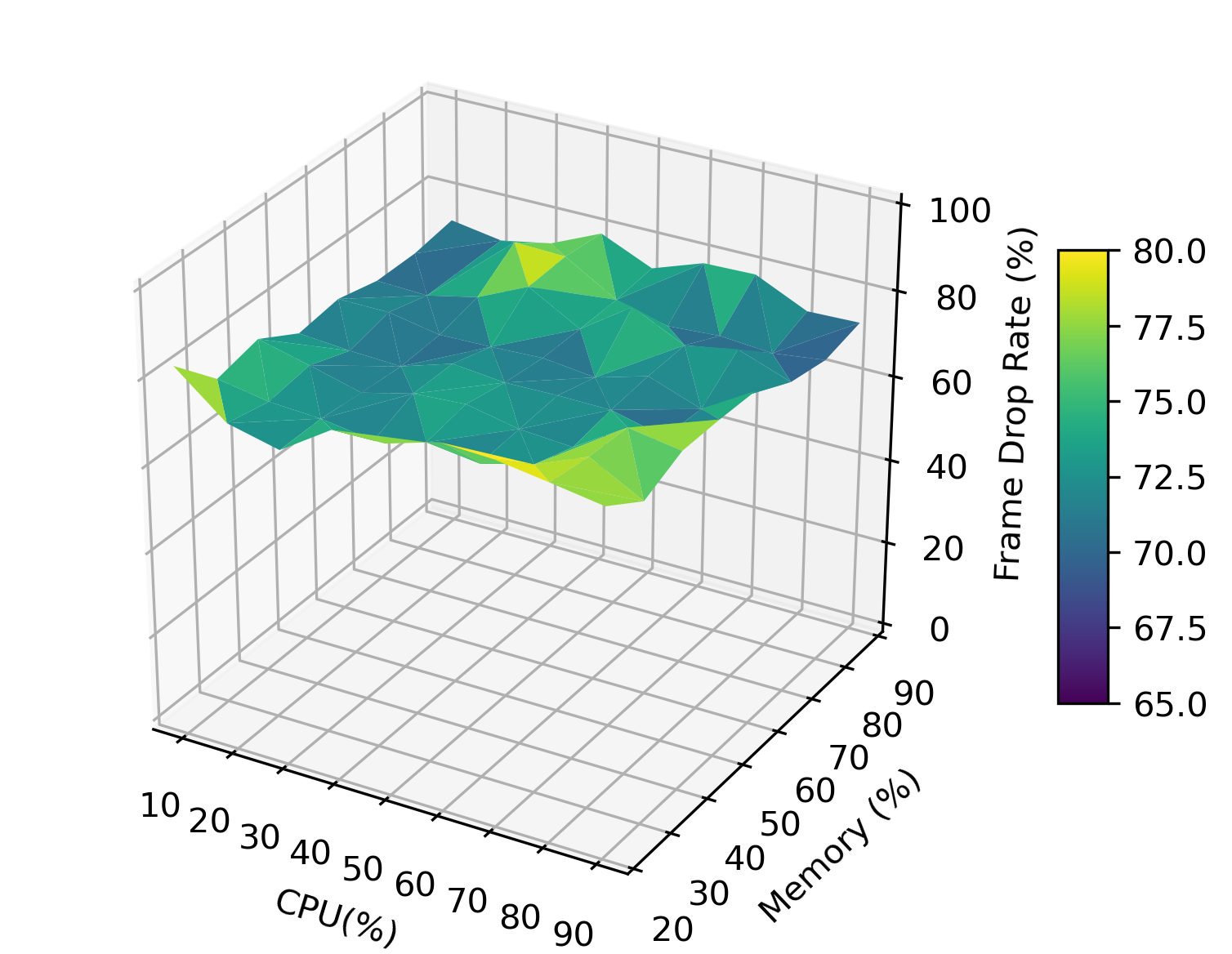}}
\end{center}
\caption{Frame drop rate during $t_{downtime}$ in the Dynamic Switching Approach for different incoming frame rates when network speed is 20Mbps}
\label{fig:framedroprate-20Mb}
\end{figure*}

\begin{figure*}[t]
\begin{center}
	\subfloat[Incoming frame rate set to 5FPS]
	{\label{fig:fps-5-5mb}
	\includegraphics[width=0.237\textwidth]
	{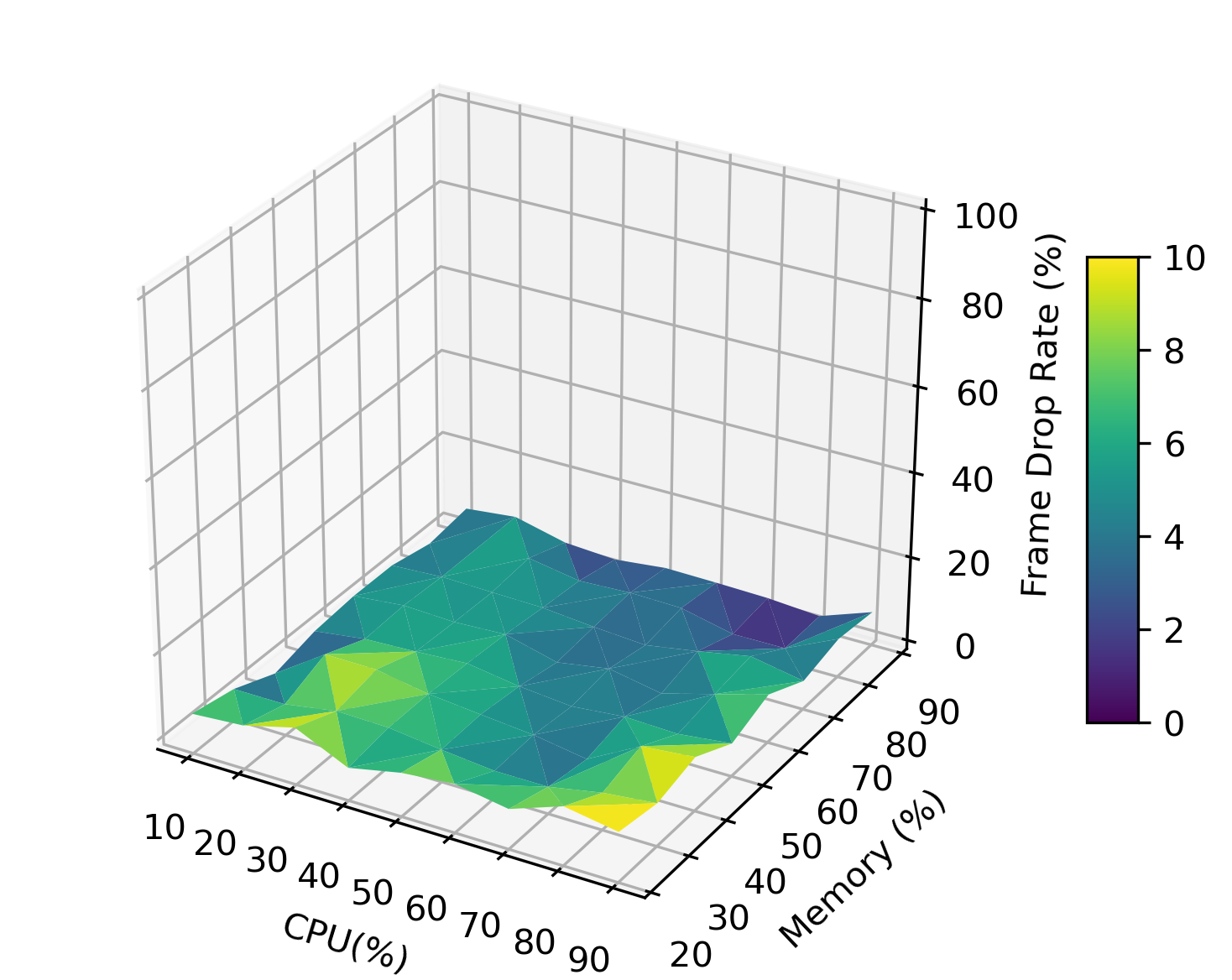}}
	\hfill
	\subfloat[Incoming frame rate set to 10FPS]
	{\label{fig:fps-10-5mb}
	\includegraphics[width=0.237\textwidth]
	{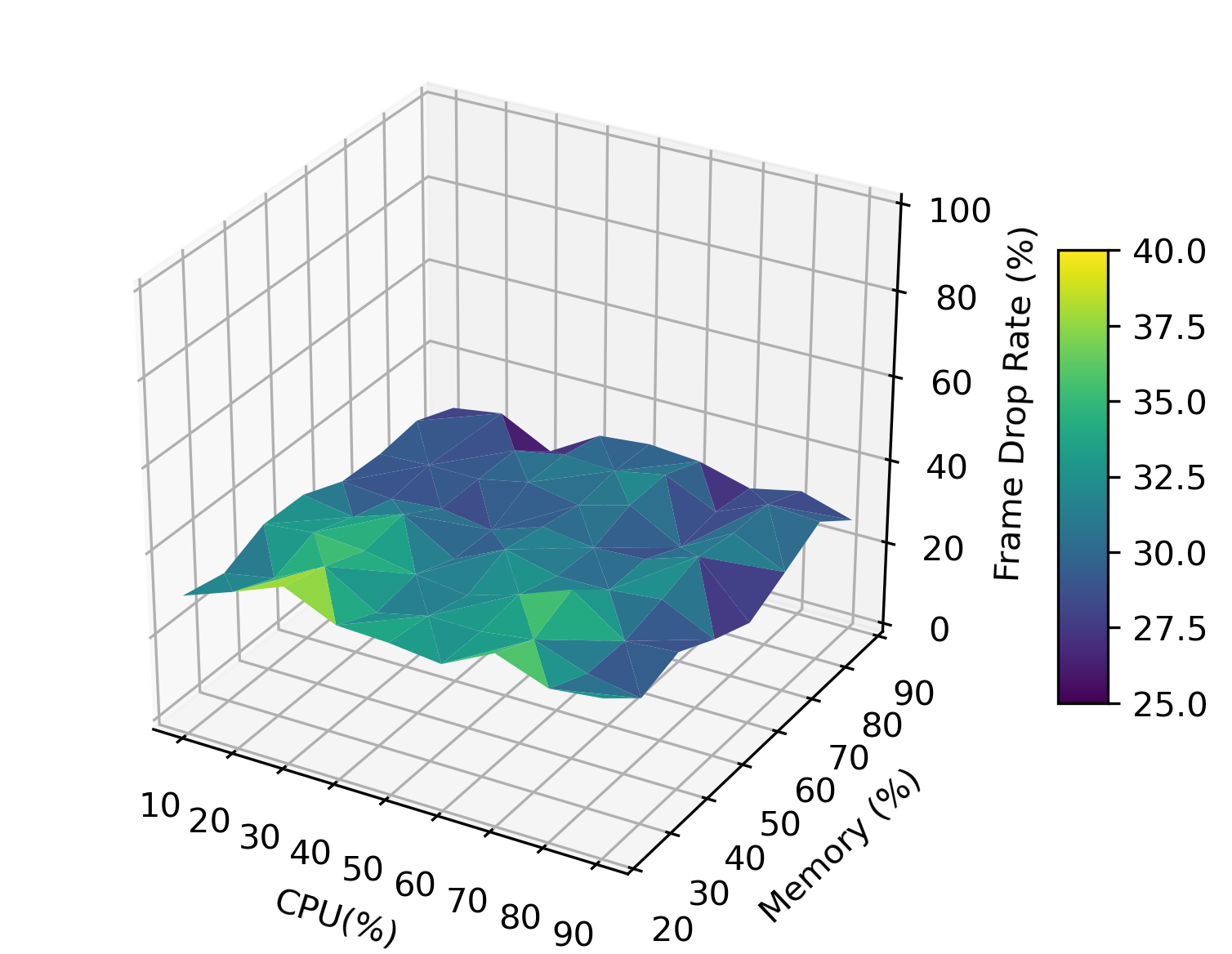}}
	\hfill
	\subfloat[Incoming frame rate set to 15FPS]
	{\label{fig:fps-15-5mb}
	\includegraphics[width=0.237\textwidth]
	{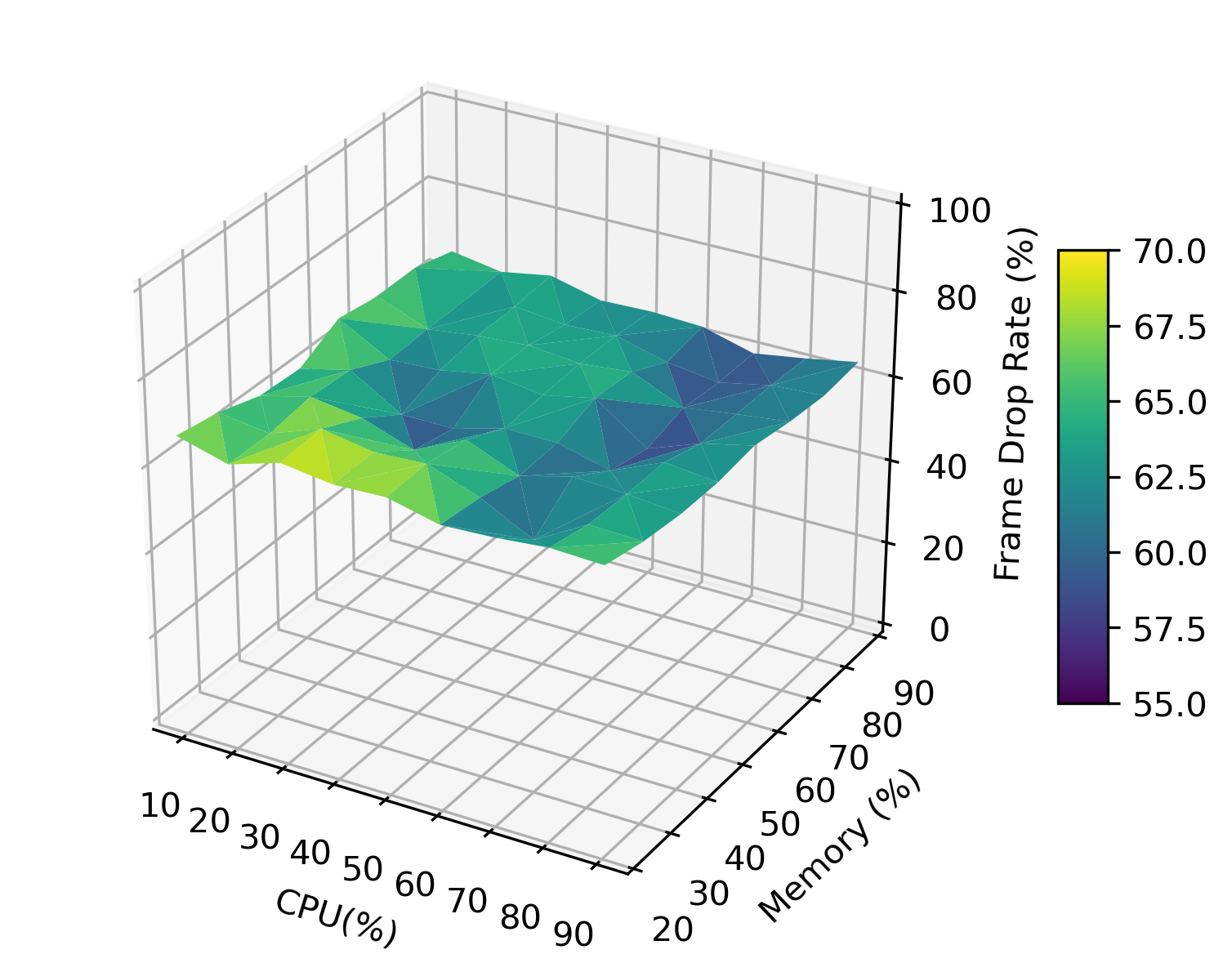}}
	\hfill
	\subfloat[Incoming frame rate set to 20FPS]
	{\label{fig:fps-20-5mb}
	\includegraphics[width=0.237\textwidth]
	{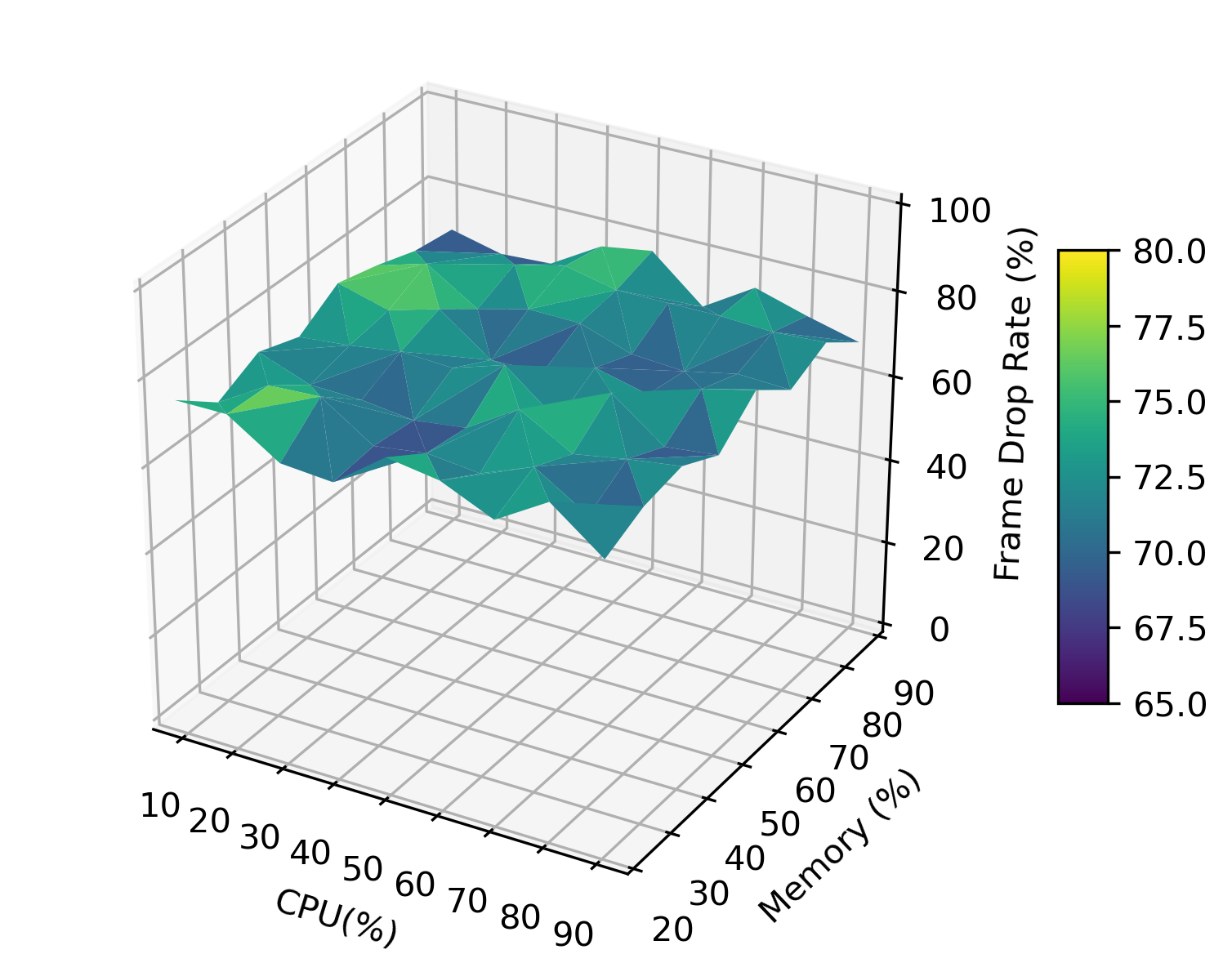}}
\end{center}
\caption{Frame drop rate during $t_{downtime}$ in the Dynamic Switching Approach for different incoming frame rates when network speed is 5Mbps}
\label{fig:framedroprate-5Mb}
\end{figure*}

Figure~\ref{fig:framedroprate-20Mb} and Figure~\ref{fig:framedroprate-5Mb} show the frame drop rate during the downtime when the network speed is 20Mbps and 5Mbps, respectively.
The x-axis denotes the percentage of CPU availability, the y-axis denotes the percentage of memory availability on edge resource, and the z-axis denotes the frame drop rate for different incoming frame rates (denoted as Frames Per Second (FPS)). The general trend is that more frames are dropped as the incoming frame rates increase.


\begin{table}[]
\caption{Total  resources  required  for  running  the  baseline and Dynamic Switching approaches in \texttt{NEUKONFIG}}
\label{tab:resources}
\begin{adjustbox}{width=.5\textwidth,center}
\begin{tabular}{|c|c|l|c|c|c|}
\hline
\multirow{2}{*}{\textbf{Approaches}} & \multicolumn{2}{c|}{\multirow{2}{*}{\textbf{Scenarios}}} & \textbf{\begin{tabular}[c]{@{}c@{}}Initial \\ Resources\end{tabular}} & \textbf{\begin{tabular}[c]{@{}c@{}}Additional \\ Resources\end{tabular}} & \textbf{\begin{tabular}[c]{@{}c@{}}Total \\ Resources\end{tabular}} \\ \cline{4-6} 
 & \multicolumn{2}{c|}{} & \textbf{\begin{tabular}[c]{@{}c@{}}Memory \\ (MB)\end{tabular}} & \textbf{\begin{tabular}[c]{@{}c@{}}Memory\\  (MB)\end{tabular}} & \textbf{\begin{tabular}[c]{@{}c@{}}Memory\\  (MB)\end{tabular}} \\ \hline
\multicolumn{1}{|c|}{\textbf{Baseline}} & \multicolumn{2}{c|}{-} & 763.1 & - & 763.1 \\ \hline
\multirow{4}{*}{\textbf{\begin{tabular}[c]{@{}c@{}}Dynamic\\ Switching\end{tabular}}} & \multirow{2}{*}{\begin{tabular}[c]{@{}c@{}}Scenario \\ A\end{tabular}} & Case 1 & 763.1 & 763.1 & 1526.2 \\ \cline{3-6} 
 &  & Case 2 & 763.1 & - & 763.1 \\ \cline{2-6} 
 & \multirow{2}{*}{\begin{tabular}[c]{@{}c@{}}Scenario \\ B\end{tabular}} & Case 1 & 763.1 & \begin{tabular}[c]{@{}c@{}}763.1\\ (during switching \\ only)\end{tabular} & \begin{tabular}[c]{@{}c@{}}1526.2\\ (of which 763.1\\  only during switching)\end{tabular} \\ \cline{3-6} 
 &  & Case 2 & 763.1 & -- & 763.1 \\ \hline
\end{tabular}
\end{adjustbox}
\end{table}

The above results highlight that the Dynamic Switching approach reduces the downtime with the added advantage that even during the downtime some frames can be processed when compared to the Pause and Resume approach. However, there is a clear trade-off between the memory resources required and the edge service downtime. This is captured in Table~\ref{tab:resources}. \textit{Initial resources} indicate the amount of resources required in memory to execute one edge-cloud pipeline. Additional resources indicates the amount of extra memory required.  
The baseline requires no additional memory as it employs the Pause and Resume technique on the same container. 

Scenario A of Dynamic Switching always has a redundant pipeline running, which has the lowest edge service downtime (less than 1 millisecond), but requires twice the amount of initial memory to keep the second pipeline running for Case 1. However in Case 2, it is noted that with the same memory requirements as the baseline approach. a redundant pipeline can be created in the same container with only a 0.6 second downtime compared to 6 seconds required by the baseline.In Case 1 of Scenario B, an additional 763.1MB is required during switching.However, in Case 2, it is noted that with the same memory requirements as the baseline approach.


The above trade-off highlights that there are multiple options to minimise edge service downtime during DNN partitioning. If additional memory is available on edge servers, then Scenario A, which has the lowest downtime can be employed. These may be suitable for larger edge resources, such as micro data centres. On the other hand, instead of using the naive approach (baseline), downtime can be reduced by an order of magnitude with the same amount of resources in Scenario A and B, Case 2. This approach captures the best of both worlds~-- reduces edge service downtime while utilising the same amount of initial memory.


\section{Related Work}
\label{sec:relatedwork}
This section presents a summary of DNN partitioning approaches and is followed by a discussion on reducing service downtime in edge computing.

\textit{DNN Partitioning:} Existing approaches for DNN partitioning at runtime aim to increase the overall performance efficiency (for example, minimising end-to-end latency) of the model by taking into account various factors, such as resource availability on the edge server or network variation.

The Adaptive Distributed DNN Inference Acceleration (ADDA) framework identifies the optimal partition point for accelerating DNN performance by considering network bandwidth and the edge server load~\cite{wang2019adda}. DNN partitions are deployed in a distributed environment using container technology~\cite{zhou2019distributing,couper}. Edgent proposes to leverage the accuracy-latency trade-off by using the early-exit strategy and DNN partitioning~\cite{li2018edge}.
The Boomerang framework determines the optimal partition point by considering an application-defined latency requirement and the available bandwidth~\cite{zeng2019boomerang}.
Research considers DNN partitioning and employs runtime parameters to improve the overall computation performance of the model~\cite{xu2019reform, dynamicDNN, zeng2020coedge, mohammed2020distributed}.

The research mentioned above has been developed to provide the initial DNN partitions, but does not account for the performance of the DNN once they are deployed. As highlighted in Section~\ref{sec:background} there is a need for considering DNN repartitioning post-deployment across the edge and cloud due to network variability. In this paper, \texttt{NEUKONFIG} identifies the scenarios when a DNN needs to be repartitioned and the optimal partitioning point for a DNN post-deployment when operational conditions change.

\textit{Edge Service Downtime:} One hypothesis that is verified in this paper is that a service downtime will be incurred at the edge for DNN partitioning using existing approaches. The classic approach employed in the literature is Pause and Resume~\cite{da2019multi}; while such an approach is not presented for DNN repartitioning it has been used for distributed streaming applications.
However, a downtime in the order of seconds is incurred by Pause and Resume for DNN repartitioning that does not make it suitable to be employed for latency-critical applications.  

Other approaches that address minimising service downtime within the context of edge computing are checkpointing, re-execution, and replication~\cite{aral2018dependency, tang2018migration, puliafito2019container}. 
The Follow-Me Chain is proposed to minimise service interruption for mobile users by formulating the service placement and migration problem using Integer Linear Programming~\cite{chen2019mobility}. To satisfy the user's QoS at runtime extra VMs are added which increases the monetary costs. To balance this trade-off, a multi-objective optimisation technique to maximise the system reliability and minimise the system cost has been proposed~\cite{yao2019fog}.

A machine learning based approach is proposed that learns the spatio-temporal dependencies between edge server failures and combines them with the topological information to incorporate link failures~\cite{aral2020learning}. Reliable services in an edge environment are considered by using a fault tolerance protocol taking into account the characteristics of the execution environment for stateful IoT applications~\cite{ozeer2018resilience}. 
A live redundancy migration scheme is proposed to reduce downtime experienced due to the stop-and-copy phase of containers~\cite{govindaraj2018container}. A dynamic online framework is proposed that configures edge clouds based on system history~\cite{hou2016asymptotically}. MOERA, which is a mobility-agnostic online algorithm is proposed to support continuous adaptation of resources in edge clouds and user mobility dynamics~\cite{wang2018moera}.

Although the above strategies have been used in classic distributed systems, they are unsuited for DNN repartitioning in (near) real-time for the edge. Methods that rely on checkpointing, replication and integer programming all have large overheads and do not attempt to minimise service downtime. Many of the above also focus on the theoretical modelling resulting in substantial gaps between theory and practice, thereby restricting their use for near real-time DNN repartitioning as envisioned in this paper.

This paper presents a practical approach (Section~\ref{sec:technique}, namely the Dynamic Switching approach integrated within the \texttt{NEUKONFIG} framework. It is the first attempt at minimising edge service downtime for DNN repartitioning.

\section{Conclusions}
\label{sec:conclusions}

Deep Neural Networks (DNNs) are distributed across the edge and the cloud to meet performance objectives. Where a DNN should be partitioned for deployment is determined based on operational conditions and when these change the DNN will need to be repartitioned and redeployed to maintain performance objectives. Existing approaches, such as Pause and Resume used for deploying a repartitioned DNN will incur a service downtime on the edge that will not be suitable for latency-critical applications requiring real-time processing. 

This paper presents an experimental framework \texttt{NEUKONFIG} that (i) identifies the scenarios in which DNN repartitioning is required, (ii) identifies the service downtime at the edge when repartitioning DNNs, and (iii) develops approaches for reducing the edge service downtime. The approaches are based on `Dynamic Switching' by making use of secondary edge-cloud pipelines to process data when repartitioning DNNs. Two scenarios are considered in the Dynamic Switching approach: (i) a second edge-cloud pipeline is always running, and (ii) a second pipeline is initialised only when the network speed changes. Experimental studies were carried out on a lab-based edge-cloud testbed to evaluate Dynamic Switching against the baseline approach based on Pause and Resume. Experimental results show that the baseline approach incurs a downtime of over 6 seconds when no data can be processed. On the other hand, when a second edge-cloud pipeline is always running, the Dynamic Switching approach has a downtime of less than 1 millisecond. However, this requires twice the amount of memory resources when compared to the baseline. In a case when an edge-cloud pipeline is initialised only when the network speed changes and requires only the same amount of memory as the baseline, the downtime is nearly 0.6 seconds. For Dynamic Switching approaches it is noted that data can be processed during the downtime albeit at a lower quality in contrast to the baseline. 

\textit{Limitations and Future Work:} Currently, \texttt{NEUKONFIG} repartitions DNN whenever there is a change in network speed which may adversely impact the performance efficiency of real-time applications. Future work will consider how frequently the DNN must be repartitioned for an application and its impact during the life-cycle of the deployed application.

\ifCLASSOPTIONcompsoc
  \section*{Acknowledgments}
\else
  \section*{Acknowledgment}
\fi
The first author was supported by a Schlumberger Foundation scholarship. The last author was supported by a Royal Society Short Industry Fellowship.
\ifCLASSOPTIONcaptionsoff
  \newpage
\fi

\bibliographystyle{IEEEtran}  
\bibliography{paper-v1}

\end{document}